\documentclass[a4paper]{spieman}  
\usepackage{amsmath,amsfonts,amssymb}
\usepackage{graphicx}
\usepackage{setspace}
\usepackage{tocloft}
\usepackage{xcolor} 
\newcommand{\micron}{\ensuremath{\mu\mathrm{m}}}
\usepackage{aas_macros} 
\usepackage{rotating} 
\title{Characterization of the GIGAPYX-4600 CMOS visible image sensor in the context of high precision astrometry}

\author[a,b,c]{Manon Lizzana}
\author[a]{Hugo Rousset}
\author[a]{Sébastien Soler}
\author[a]{Fabrice Pancher}
\author[d]{\'Eric Thi\'ebaut}
\author[a]{Fabien Malbet}

\affil[a]{Univ.\ Grenoble Alpes, CNRS, IPAG, F-38000 Grenoble, France}
\affil[b]{CNES, F-75001 Paris, France}
\affil[c]{Pyxalis, F-38430 Moirans, France}
\affil[d]{Université Lyon 1, ENS de Lyon, CNRS, CRAL, UMR 5574, Saint-Genis-Laval, France}

\cftpagenumbersoff{figure}
\cftpagenumbersoff{table} 
\begin{document} 
\maketitle

\begin{abstract}
To evaluate the potential of detectors for next-generation space astrometry missions such as the HWO and Theia, an extensive analysis of the performance of the GIGAPYX-4600 46-megapixel CMOS detector from Pyxalis was conducted at IPAG. The $5456 \times 8320$-pixel detector, featuring a $4.4$-micron pitch and back-side illumination, constitutes a matrix of $24 \times 36$\,mm. It has a measured gain of $13.3\pm0.9\,\mathrm{e}^{-}/\mathrm{ADU}$ in low-gain mode. The saturation level of the detector in low-gain-mode is estimated to be approximately 52\,ke$^{-}$. This sensor demonstrates linearity errors of $0.90 \pm 0.18\%$, pixel response non-uniformity $\leq1.3\%$, and readout and quantization noise of $12\pm3$\,e$^-$ in low-gain mode. Dark current was measured at $35\pm1$\,e$^-$/s at $35^\circ$C, defective pixels accounted for only $8\times10^{-5}\%$ of the array, and inter-pixel capacitance remained under 3\%. We demonstrated that inter-pixel capacitance effects can be effectively mitigated using spatially varying kernel correction algorithms. Radiation tests confirmed its robustness against protons and heavy ions, and no latch-up was observed. These measurements agree with the manufacturer figures. These characterization results validate the GIGAPYX-4600 as a strong candidate for high-precision astrometry. Further improvements are currently being developed for space implementation.
\end{abstract} 

\keywords{characterization, detectors, astrometry, Theia, HWO}

{\noindent \footnotesize\textbf{*}Manon Lizzana,  \linkable{manon.lizzana@univ-grenoble-alpes.fr}}


\section{Introduction}
\label{part:intro}

\begin{figure}
    \centering
    \includegraphics[width=0.8\linewidth]{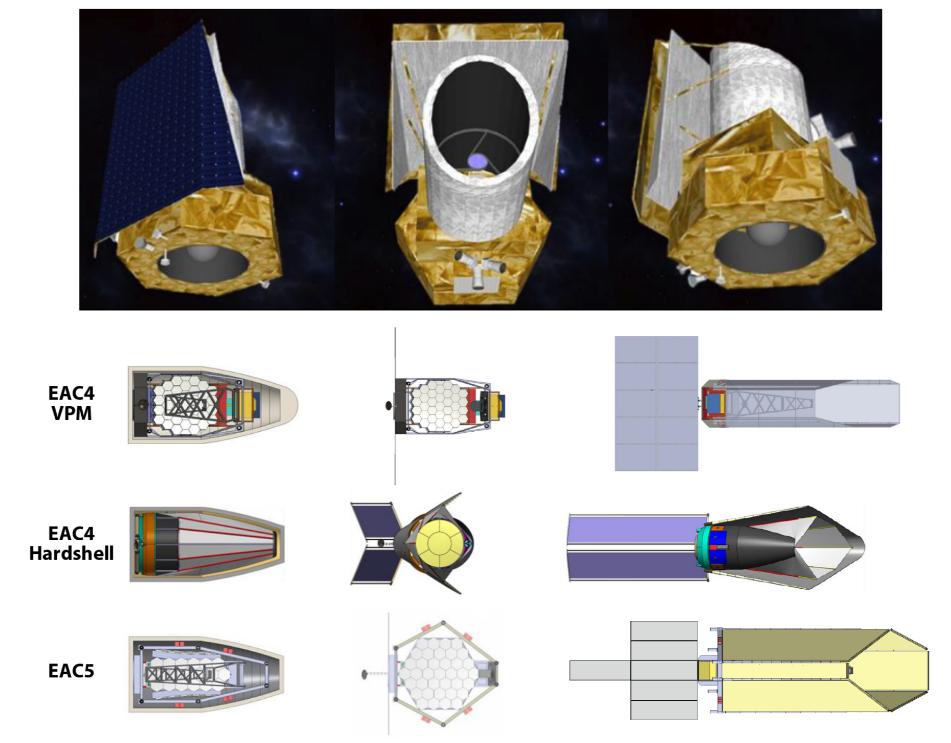}
    \caption{Top: different views of the proposed Theia satellite concept \cite{Malbet_2021}. Bottom: different concepts proposed for the HWO satellite \cite{Feinberg_2026}. The left column shows the stowed launch configuration, the middle column is looking down the telescope bore sight, and the right column shows the top-down view.}
    \label{fig:telescopes_Theia_HWO}
\end{figure}

Astrometry evaluates the position, proper motion, and parallax of celestial objects. Current research, particularly in exoplanet science, demands exceptionally high accuracy, below 0.3 microarcseconds~\cite{Malbet_2016,Malbet_2025}. Such precision could be achieved by future space missions, like NASA’s next flagship mission, Habitable Worlds Observatory~\cite{Feinberg_2024} (HWO), and, the proposed Theia mission~\cite{Malbet_2021} submitted to the European Space Agency (ESA). HWO features a 7-meter diameter mirror and will potentially incorporate an astrometric mode~\cite{Malbet_2026} as part of the High Resolution Imager (HRI) instrument, while Theia~\cite{Malbet_2022} is a dedicated astrometry 0.8-m telescope project aimed at achieving this unprecedented accuracy in celestial measurements. Both missions operate in the visible domain. The two telescopes are illustrated in Fig.~\ref{fig:telescopes_Theia_HWO}.

The HRI’s Field of View (FOV) is projected to cover approximately $3\times 4\,\mathrm{arcmin^2}$ with a focal length of $130$~m, requiring roughly $22,000\times30,000$ pixels, or $7\cdot10^8\,\mathrm{px}$ of $4.4\,\micron$. Similarly, Theia’s FOV is planned to span $30\times 30\,\mathrm{arcmin^2}$ with a $13$~m focal length, necessitating a detector of approximately $25,000\times25,000$ pixels, or $6\cdot10^8\,\mathrm{px}$. In both cases, the scale approaches $\sim 1$\,Gpx. To address this issue, we propose to provide a focal plane assembly composed of detectors~\cite{Amiaux_2025} with number of pixels between 200 and 300\,Mpx.

The GIGAPYX-4600 CMOS detector, manufactured by Pyxalis (Moirans, France) is part of the Gigapyx series, which ranges from 14\,Mpx to 220\,Mpx. This detector represents a compelling proof of concept for high-precision astrometry applications since it is for the moment the only detector available for testing, while four Gigapyx 220MP detector could form the architecture of astrometric focal planes~\cite{Amiaux_2025}. 

Section~\ref{sec:description} of this paper provides a detailed description of the GIGAPIX-4600 CMOS sensor, as advertised by its manufacturer, Pyxalis. Section~\ref{sec:methodology} presents the methodology for characterizing the sensor, and Sec.~\ref{sec:results} details the results obtained at IPAG (Institut de planétologie et d'astrophysique de Grenoble) in Grenoble, France. Section~\ref{sec:discussion} discusses the performance of this detector in the context of a high-precision astrometric instrument.

\section{Detailed description of the GIGAPYX-4600}

\subsection{Technical description of the GIGAPYX-4600 CMOS sensor}
\label{sec:description}

\begin{figure}[t]
    \centering
    \includegraphics[width=0.9\linewidth]{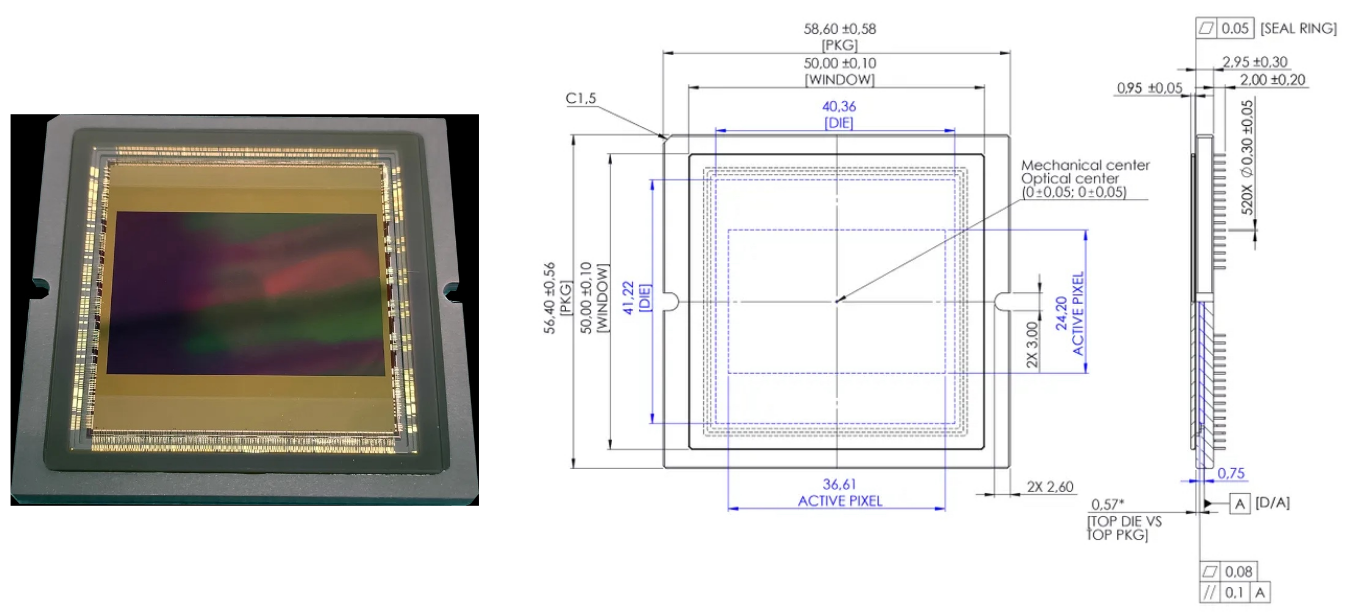}
    \caption{GIGAPYX-4600 : picture of the detector (left) and mechanical drawing (right) provided by Pyxalis product page (the lengths are in mm).}
    \label{fig:Gigapyx4600}
\end{figure}

\begin{table}
    \centering
    \caption{Summary of the features of the GIGAPYX-4600 sensor by Pyxalis$^*$}
    \medskip
    \label{tab:Gigapyx4600}
    \begin{tabular}{ll}
    \hline  
    \textbf{Features} &\textbf{Values}\\
    \hline
    \hline
    Shutter modes &Rolling Shutter\\
    Resolution &46 MegaPixels\\
    Active Pixels &$8320 \times 5456$\\
    Image Size &$24 \times 34$\,mm\\
    Aspect Ratio &$3:2$\\
    Back-Side Illumination (BSI) Technology &Yes\\
    Pixel Size &4.4\,\micron\\
    Frame rate &150\,fps in Full Frame, 200f\,ps in 8k format\\
    Ultra-Low Crosstalk &Deep Trench Isolation Technology\\
    High Dynamic Range &Up to $90$~db in Dual Gain\\
    Packaging &Ceramic package, \textmu{}PGA, 520 pin\\
    Power &$<4.1$~W at full speed\\
    Color Filter Array &Monochrome without microlenses\\
    Features (ROI, Parameters ...) &6 $\times$ Contexts\\
    Output Format &8 to 128 x LVDS/Sub-LVDS\\
    SPI interface controller &Yes\\
    I/O &For Trigger, Shutter \& Flash synchronization\\
    \hline
    {\small $^*$\href{https://pyxalis.com/products/gigapyx-4600/}{https://pyxalis.com/products/gigapyx-4600/}}&\\
    \end{tabular}
\end{table}

The GIGAPYX-4600 is a CMOS image sensor (Fig.~\ref{fig:Gigapyx4600}) at the forefront of technology designed to meet the rigorous demands of high-performance imaging applications whose features are described in Table~\ref{tab:Gigapyx4600}. As the first release in the Gigapyx family, it delivers a 46 megapixels of resolution in a full-frame 35-millimeters optical format. This makes it suited for professional cinematography, scientific imaging, aerial and space applications, and high-resolution industrial inspection. Its back-side illumination (BSI) architecture, combined with a $4.4\,\micron$ pixel pitch, ensures optimal light sensitivity and low noise, enabling high-fidelity image capture even in challenging lighting conditions.

The QE of the GIGAPYX-4600 spans the entire visible spectrum ($\sim 300-800$\,nm) and and, to a lesser extent, up to $1100$\,nm. The GIGAPYX-4600 operates in rolling-shutter mode with three distinct modes: high-gain, low-gain, and in-pixel true high dynamic range (HDR). The results presented in this paper were obtained in low-gain mode because this mode will be used in astrophysical applications to collect the same amount of light with a smaller number of images. HDR mode enables the sensor to simultaneously capture intricate details in bright and dark areas of a scene, making it ideal for applications requiring intra-scene dynamic range such as astronomy. Although these characteristics may be of interest, this mode of operation has not been studied in this paper. The sensor offers full wells of 5,000 electrons (high-gain mode) and 50,000 electrons (low-gain mode).  It ensures a high image quality and a high accessible dynamic range. Correlated double sampling (CDS) is also implemented to significantly reduce fixed pattern and temporal noise. The sensor features ultra-low readout noise of 1.5\,e$^-$ in high-gain mode.

The GIGAPYX-4600 supports high-speed imaging, with a maximum frame rates of 150 frames-per-second (fps) in full-frame mode and 200\,fps in the 8K format ($8320\times4320$~px). The deep trench isolation technology, minimizes carrier diffusion from one pixel to its neighbors and enhances image clarity~\cite{Ahmed_2014}. The sensor is available in both monochrome and standard RGB Bayer configurations, offering flexibility for various imaging needs. The monochrome version was used in this study because it is the one required for astrometry applications. It also features eight to 128 LVDS data lanes for high-speed data transfer and consumes less than 4.1W of power at full speed. This makes the GIGAPYX-4600 efficient and versatile for integration into advanced imaging systems.

The sensor's ceramic \textmu{}PGA package and 520-pin interface, ensures robustness and reliability, providing consistent performance in demanding environments. Pyxalis provides an evaluation kit, called GENEPYX which consists in a camera and its electronics, to test the sensor’s capabilities and facilitate its adoption in cutting-edge imaging applications. To evaluate the astrometric capability of the GIGAPYX-4600 sensor, we operate it with the GENEPYX camera demo kit.

\subsection{Detector specifications}

To study Sun-like stars, the detector's quantum efficiency must be distributed across the visible spectrum. Moreover, a Nyquist sampling is required to reconstruct the PSF. In other words, the pixel size $e$ (the sampling spatial period) must be $2e\leq f\lambda/D$, with $\lambda$ the wavelength and $D$ the telescope aperture. In the context of the Theia mission, considering $f=13$~m and $D=0.8$~m , the pixel size should be $e_{\mathrm{Theia}}<4.5~\mu$m. In the same way, in the context of the HWO mission, considering $f=130$~m and $D=7$~m, the pixel size should be $e_{\mathrm{HWO}}<5.1~\mu$m. In addition, as seen in Sec.~\ref{part:intro}, the giga-pixel resolution is required to cover the entire FOV. The brightest stars require an exposure time of $2$~ms in the Theia configuration and $30~\mu$s in the HWO configuration. Because they are very bright, the detector must be able to read the signal after this exposure time, however the frame rate is relaxed because it allows photons to be lost between two frames. Less bright stars require an exposure time of $0.4$~s in the Theia configuration and $2$~s in the HWO configuration. To avoid losing photons, the detector must be able to read at this frame rate. Finally, there must be no microlenses to avoid unmeasurable astrometric effects.

In summary, the detector must be sensitive across the entire visible spectrum. For Nyquist sampling, a pixel size of $4$ or $5~\mu$m is required. The resolution must enable a giga-pixel focal plane to be covered by a small number of detectors (between 1 and 8, for example, if they cannot be connected end-to-end to avoid dead zones). The frame rate must be compatible with an exposure time of $2$~ms for Theia and $30~\mu$s for HWO.

\subsection{Comparison between the GIGAPYX-4600 and other detectors}

Table~\ref{tab:tableau_comparaison_détecteurs} compares the characteristics of different detectors that could be used for high-precision astrometry. The table compares detectors from five companies: Canon (Japan), Teledyne (USA), Sony (Japan), Gpixel (China), and Pyxalis (France). These detectors have several features in common, such as large format, back illumination, and rolling shutter readout.

\begin{sidewaystable}
    \caption{Comparison of detectors that could be used for high-precision astrometry}
    \centering
    \begin{tabular}{c|c|c|c|c|c|c|c}
     company & detector name & type & resolution (HxV) & nb px (Mpx) & px size ($\mu$m) & ship size (mm) & package size (mm) \\
    \hline
    Canon (Japan) & 120MXS & CMOS & $13272 \times 9176$ & $122$ & $2.2$ & $29.22 \times 20.20$ & - \\
    Canon (Japan) & LI8020SA & CMOS & $19568 \times 12588$ & $250$ & $1.5$ & $29.35 \times 18.88$ & - \\
    Canon (Japan) & LI8030SA & CMOS & $24592 \times 16704$ & $410$ & $1.5$ & diagonal $44.6$ & $51.24 \times 37.42$ \\
    Teledyne (USA) & CCD290-99 & CCD & $9216 \times 9232$ & $85$ & $10$ & $92.2 \times 92.4$ & $98.5 \times 93.7$ \\
    Teledyne (USA) & CCD231-C6 & CCD & $6144 \times 6160$ & $38$ & $15$ & $92.2 \times 92.4$ & $98.5 \times 93.7$ \\
    Sony (Japan) & IMX411ALR & CMOS & $14208 \times 10656$ & $151$ & $3.76$ & diagonal $66.7$ & $60.30 \times 47.90$ \\
    Sony (Japan) & IMX811-AAMR & CMOS & $19240 \times 12840$ & $247$ & $2.81$ & diagonal $64.84$ & $75.0 \times 63.6$ \\
    Sony (Japan) & IMX661-AAMR & CMOS & $13472 \times 9568$  & $129$ & $3.45$ & diagonal $56.73$ & $75.0 \times 63.6$ \\
    Gpixel (China) & GSENSE 1081BSI & CMOS & $8900 \times 9120$ & $81$ & $10$ & $89 \times 91$ & $92 \times 102$ \\
    Pyxalis (France) & Gigapyx $220$ & CMOS & $16640 \times 13156$ & $220$ & $4.4$ & $73 \times 58$ & $77 \times 75$ \\
    \end{tabular}
    \newline

    \begin{tabular}{c|c|c|c|c|c|c}
    detector name & readout noise (e$^-$) & max readout rate (fps) & well (ke$^-$) & dark current & microlenses & space qualification \\
    \hline
    120MXS & - & $9.4$ & - & $2.3$ e$^-$ (room temperature) & No a priori & No \\
    LI8020SA & - & $5$ & - & $3.8$ e$^-$ & No a priori & No \\
    LI8030SA & - & $8$ & $3900$ & $6$ e$^-$/s (60$^{\circ}$C) & No a priori & No \\
    CCD290-99 & 4 at 0.5 MHz, 2.5 at 50 kHz & 3M & $90$ & $4$ e$^-$/h (-100$^{\circ}$C) & No a priori & - \\
    CCD231-C6 & 5 at 1MHz, 2 at 50kHz & 3M & 350 & 3 e$^-$/h (-100$^{\circ}$C) & No a priori & - \\
    IMX411ALR & - & $6$ & - & - & Yes & No \\
    IMX811-AAMR & - & $12$ & - & - & Yes & No \\
    IMX661-AAMR &  - & $21$ & - & - & Yes & No \\
    GSENSE 1081BSI & $5.35$ & $0.34$ & $91$ & $0.00373$ e$^-$/s (-70$^{\circ}$) & No a priori & - \\
    Gigapyx $220$ & $16$ & $20$ & $50$ & $34$ e$^-$/s (35$^{\circ}$C) & No & Yes \\
    \end{tabular}
    \label{tab:tableau_comparaison_détecteurs}
\end{sidewaystable}

Canon proposes detectors with small pixels, then $2\times2$ or $3\times3$ binning would be necessary, thus 2 or 3 gigapixels would be mandatory rather than 1. The focal plane would be composed of many detectors. Moreover, they have never been tested in a space environment. Teledyne manufactures relatively small formats (for example, around 12 CCD290-99 detectors would be necessary to reach gigapixel resolution). Since they cannot be perfectly connected end-to-end, FOV will have a sampled area and dead zones. Additionally, the pixel size is larger than desired, which leads to undersampling. The detectors proposed by Sony are interesting, but they have a micro-lens matrix to improve efficiency. These elements disrupt the spatial response of the pixels and can artificially shift their centroid, which is incompatible with high-precision astrometry. Furthermore, they have never been tested in a space environment. The problems encountered with the Gpixels detectors are the same as those we faced with Teledyne: small formats and large pixels.

\section{Characterization Methodology}
\label{sec:methodology}

In this section, we detail the experimental setup and the image acquisition sequence that we used to achieve to fully characterize the GIGAPYX-4600 detector.

\subsection{Experimental Setup}

\begin{figure}
    \centering
    \includegraphics[width=\linewidth]{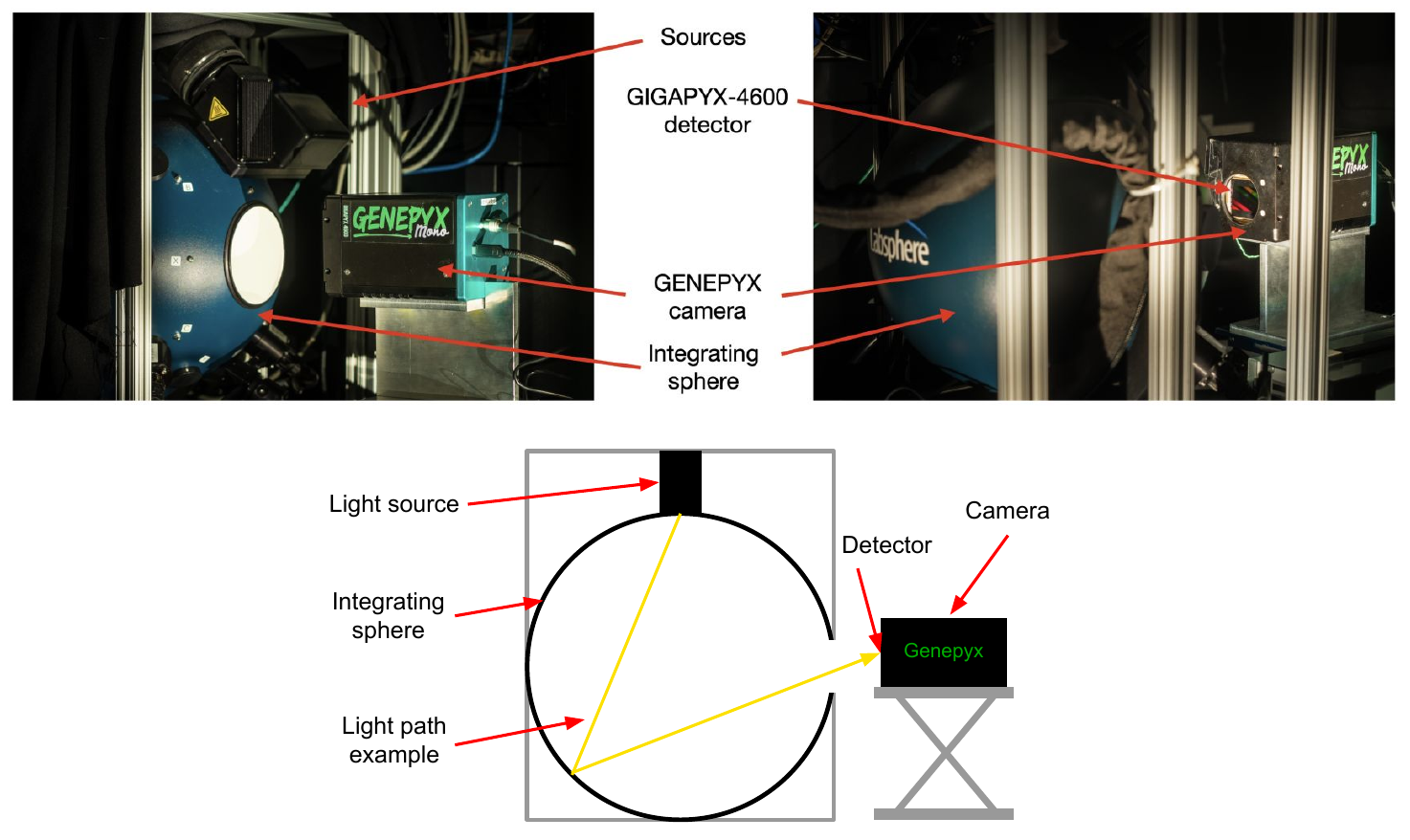}
    \caption{Photo and schematic of the characterization testbed. The Heliosplus USLR-D12F-NDNN-P from Labsphere placed in front of the evaluation kit camera GENEPYX with the GIGAPYX-4600 sensor.}
    \label{fig:testbed}
\end{figure}

We characterized the A1 version of the GIGAPYX-4600 detector using the GENEPYX camera manufactured by Pyxalis. The signal is encoded in 12 bits, or 4,096 analog-to-digital units (ADUs). The camera is controlled by a computer using the Pyxalis-developed SDK (software development kit). The characterization testbed, developed at IPAG, is shown in Fig.~\ref{fig:testbed}. The detector is placed in front of an integrating sphere which provides a flat field. The HeliosPlus USLR-D12F-NDNN-P integrating sphere from Labsphere was used, it has an output diameter of 10\,cm and a spatial luminance uniformity better than 2\%.

\subsection{Image Acquisition}

All the characterization steps follow the EMVA~1288 standard (release 4.0)\cite{EMVA} for characterizing image sensors and cameras. Accurate CMOS imaging typically requires a set of bias frames, dark frames, and flat field frames that are used to remove the individual pixel bias pattern, dark current pattern, and detector response pattern. In the following, $\mu_{X}$ and $\sigma^2_{X}$ will refer respectively to the spatial mean value and the variance of the signal $X$ where $X$ can be $B$ for the bias measurements, $D$ for the dark ones and $F$ for the flat ones. For each pixel, the signal value used is the median of the set of temporal signal values. We checked that the pixel signal follows a Gaussian statistics except for some defective pixels, the median and mean yield the same value, but the median is more robust against outliers. The spatial mean values and the variances are computed without the defective pixels identified in Sec.~\ref{sec:defect_px}. To evaluate the detector's behavior, we captured these three types of images :
\begin{itemize}
    \item bias images: acquisition without light (camera closed by a cap) during a negligible exposure time ($t_{\mathrm{exp}}=60\,\mu$s)
    \item dark images: acquisition without light (camera closed by a cap) during a non-negligible exposure time ($25$\,ms, $10$\,s, $20$\,s, $30$\,s, $40$\,s, $50$\,s, and $60$\,s)
    \item flat images: acquisition using the integrating sphere during a non-negligible exposure time (20\,$t_{\mathrm{exp}}$ from $2$ to $47$\,ms)
\end{itemize}

\subsection{Discussion about the temperature}

\begin{figure}
    \centering
    \includegraphics[width=0.4\linewidth]{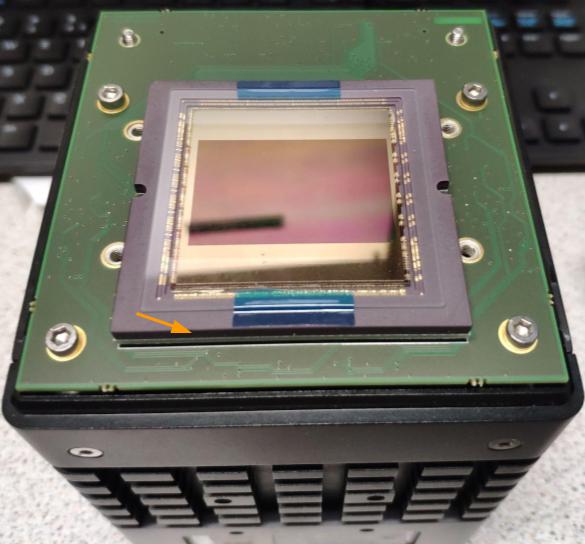}
    \caption{Orange arrow indicates the place were the type K thermocouple were installed to measure the detector temperature.}
    \label{fig:Ggapyx_sondeT}
\end{figure}

Some results can be very sensitive to temperature variations, especially those concerning dark current measurements. We installed two temperature sensors: one on the ceramic next to the detection matrix and one next to the camera to measure the room temperature. The sensors used are type K thermocouple, the one on the ceramic was placed as indicated in Fig.\ref{fig:Ggapyx_sondeT}. The goal is not to measure the detector's exact temperature, but rather to study its variations over time.

Image acquisitions are taken at $35^\circ$C, in a laboratory room with a regulated temperature of approximately $\sim20^\circ$C. The characterization was conducted at $35^\circ$C because that corresponds to the stabilization temperature when the detector is kept at room temperature. No temperature control system was available for our setup. The detector temperature is quite stable, varying by less than $0.1^\circ$C.

The data were collected following a stabilization period ($\sim1$~h) during which the power consumption was stabilized by running the detector in streaming mode and discarding the data. The operating temperature at equilibrium was then reached. A more precise study will be provide in the context of dark measurements in Sec.~\ref{sec:dark_current}

\section{Results}
\label{sec:results}

\subsection{Master bias, dark and flat images}
\label{sec:masters}

A total of 170 images are collected to create the flat fields used to estimate the linearity and to compute the photon transfer curve (PTC), presented respectively in Sec.~\ref{sec:PTC} and Sec.~\ref{sec:Response_curve}.
The median and variances are computed using 300 images, for the readout noise presented in Sec.~\ref{sec:roq_noise}, and for the pixel response non-uniformity (PRNU) presented in Sec.~\ref{sec:PRNU}. A total of 50 images are used for the dark current calculation presented in Sec.~\ref{sec:dark_current}. 

The "master flat", "master dark" and "master bias" are computed as the median over the image cubes for each exposure time. The darks are corrected for the master bias and the flats are corrected for the master dark and the master bias.

The given uncertainties are computed as the standard deviation over the detector. When applicable, the ADU-to-electrons conversions use the gain established in Sec.~\ref{sec:PTC}.

\begin{figure}
    \centering
    \includegraphics[width=0.6\linewidth]{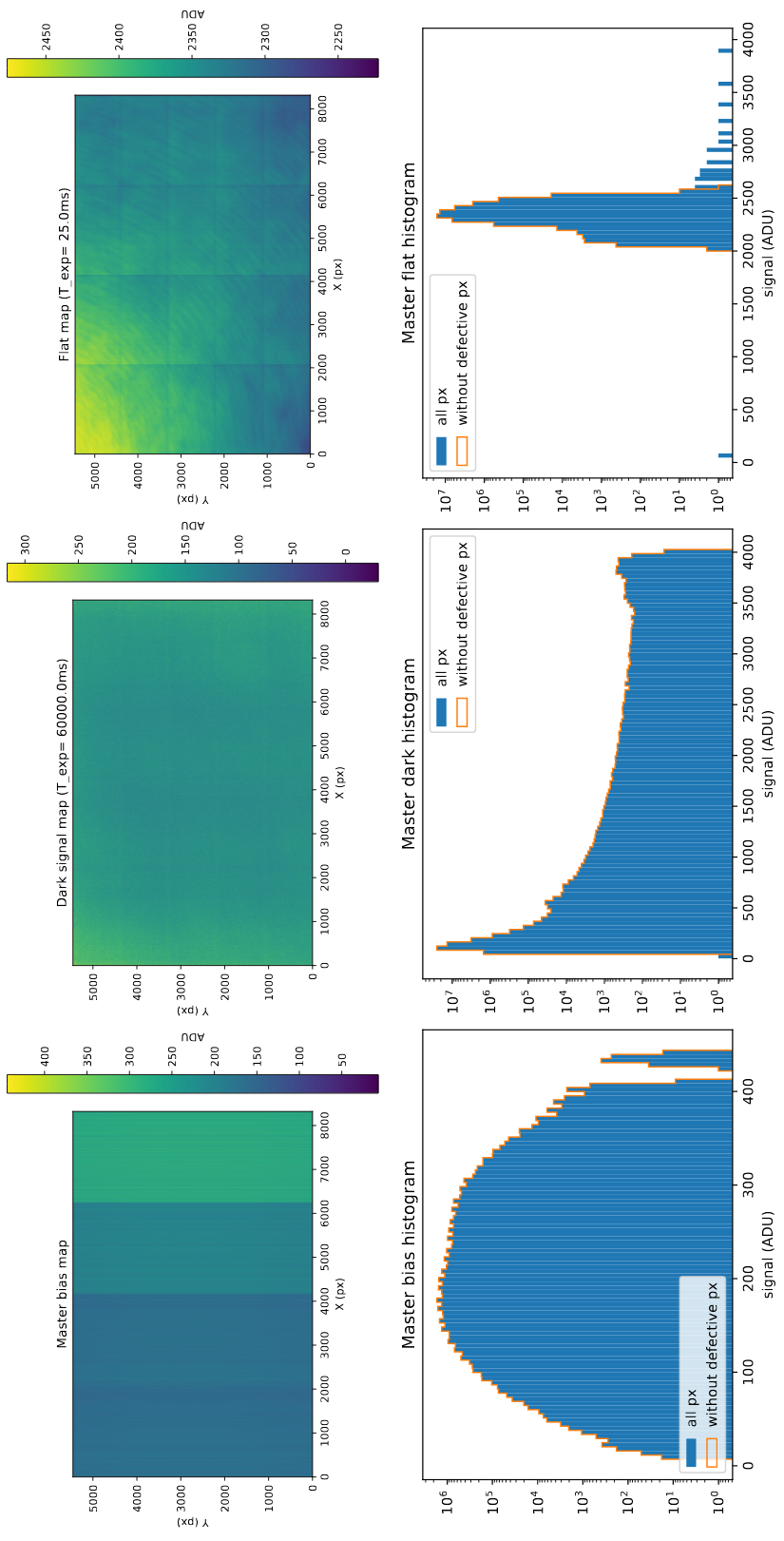}
    \caption{Master images used to characterize the detector. Top left: master bias; top center: master dark, top right: master flat. Bottom: Examples of the bias, dark and flat histograms with and without the defective pixels. In orange the identified defective pixels and in blue all the pixels (defective + functional). See part~\ref{sec:defect_px} for the defective pixels identification.}
    \label{fig:acquired_images}
\end{figure}

Examples of master bias, dark and flat images, as well as the corresponding histograms, are shown in Fig.~\ref{fig:acquired_images}. As it can be seen with the figures, the bias is structured in columns because the pixels are read in columns, and the analog-to-digital converters (ADCs) are located at the bottom and top of each column. There are 4 vertical and 5 horizontal parts, which correspond to the stitch lines of the manufacturing process. A large-scale pattern is visible in the dark map, it is probably due to the uneven temperature of the surface detector caused by the dissipation of power from the proximity electronics on the sides. Heterogeneity is visible in the flat images (the top left corner has higher values, while the other corners are darker), as well as circular patterns looking like "tree rings", which are not due to the integrating sphere. According to the manufacturer, these patterns are likely to be caused by the thinning and conditioning of the back-side interface of the detection layer.

\subsection{Photon Transfer Curve (PTC), Gain, and Dynamic Range}
\label{sec:PTC}

\begin{figure}
    \centering
    \includegraphics[width=0.6\linewidth]{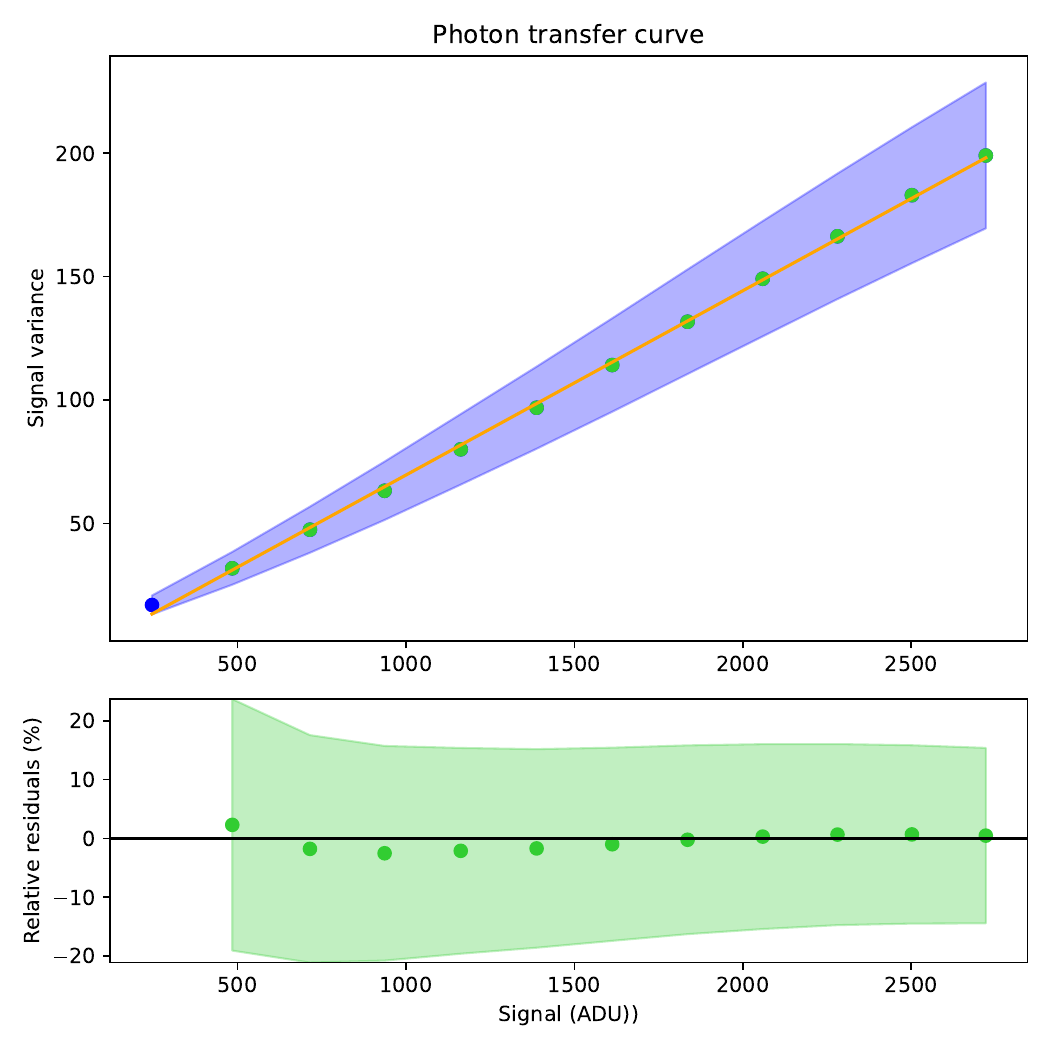}
    \caption{The photon transfer curve is plotted at the top: $\sigma_F$ as a function of $\mu_F$}. Mean signal (green dots), standard deviation on the whole detector pixels (blue area) and linear model (orange) are displayed in the top panel. The relative residuals are plotted in the bottom panel.
    \label{fig:PTC}
\end{figure}

In the photon noise regime, we have $\sigma_F^2\,[\mathrm{e}^-]=\mu_F\,[\mathrm{e}^-]$ where $\sigma_F^2$ is the temporal variance of the signal $F$ and $\mu_F$ is the mean value. We define the gain (or $K$-factor) $K$ as the conversion factor between electrons and ADU (in ADU/e$^-$): $\mu_F\,[\mathrm{ADU}]=K \mu_F\,[\mathrm{e}^-]$. The PTC~\cite{Janesick_2007} gives per pixel the temporal variance on a set of acquisitions $\sigma_F^2\,[\mathrm{ADU}]$ as a function of the signal mean value $\mu_F\,[\mathrm{ADU}]$. The PTC is plotted in Fig.~\ref{fig:PTC}. This curve is linear if the conversion factor is constant with the pixel value and the slope represents $K$ of each pixel (in ADU/e$^-$):
\begin{equation}
    \sigma_F^2=K\,\mu_F.
\end{equation}

The saturation signal value is estimated to be just above the point at which $0.1\%$ or $0.2\%$ of the total number of pixels show the maximum value. In our configuration, this corresponds to an exposure time of $39.5$~ms and a mean flat value of $3773$~ADU (according to the response curve in Fig.\ref{fig:Response_curve}). The slope of the PTC (in ADU/e-) is fitted from $0$ to $70\%$ of the saturation.

\begin{figure}
    \centering
    \includegraphics[width=0.9\linewidth]{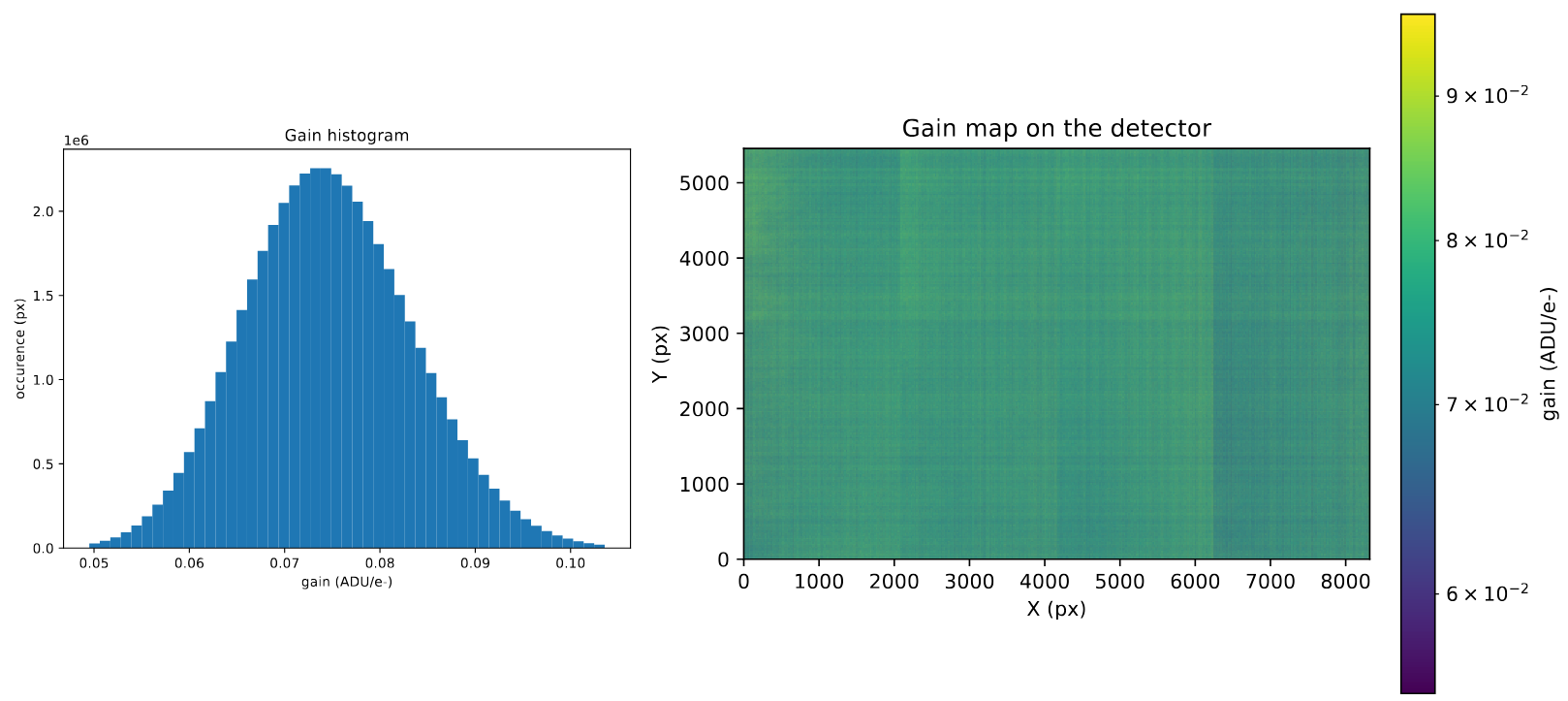}
    \caption{Map and histogram of the gain over the detector.}
    \label{fig:gain}
\end{figure}

The gain over the detector is displayed in Fig.~\ref{fig:gain}. The histogram shows good Gaussian statistics, and the map highlights a row/column structure, indicating that the gain heterogeneity most likely originates from the detector's readout circuitry. The mean gain is $K=0.075\pm0.005$\,ADU/e$^-$, corresponding to a conversion ADU-to-electrons $q = 1/K$ of $13.3\pm0.9$\,e$^-$/ADU. As shown in Fig.~\ref{fig:PTC}, the residuals indicate that the gain remains relatively constant with respect to the pixel value. This behavior was expected by the manufacturer.

As will be discussed in Sec.~\ref{sec:IPC}, our detector is subject to inter-pixel capacitance (IPC). However, Poisson noise can be attenuated, resulting in an underestimation of the variance and, consequently, the gain~\cite{Moore_2006}. The gain expressed here was not corrected for IPC.

We measured a bias of $195$~ADU, and the detector provides a 12-bit signal. Therefore, the useful dynamic range is  $4096-195=3901$~ADU . As previously demonstrated, the measured gain is $K=0.075\pm0.005$\,ADU/e$^-$; therefore, the saturation charge (or useful full well capacity) is estimated to be $3901/0.075\simeq52$\,ke$^-$. However, this does not represent the actual full well capacity of the detection layer, but rather the saturation of the ADC in our configuration.

\subsection{Response Curve and Linearity Error}
\label{sec:Response_curve}

\begin{figure}
    \centering
    \includegraphics[width=0.6\linewidth]{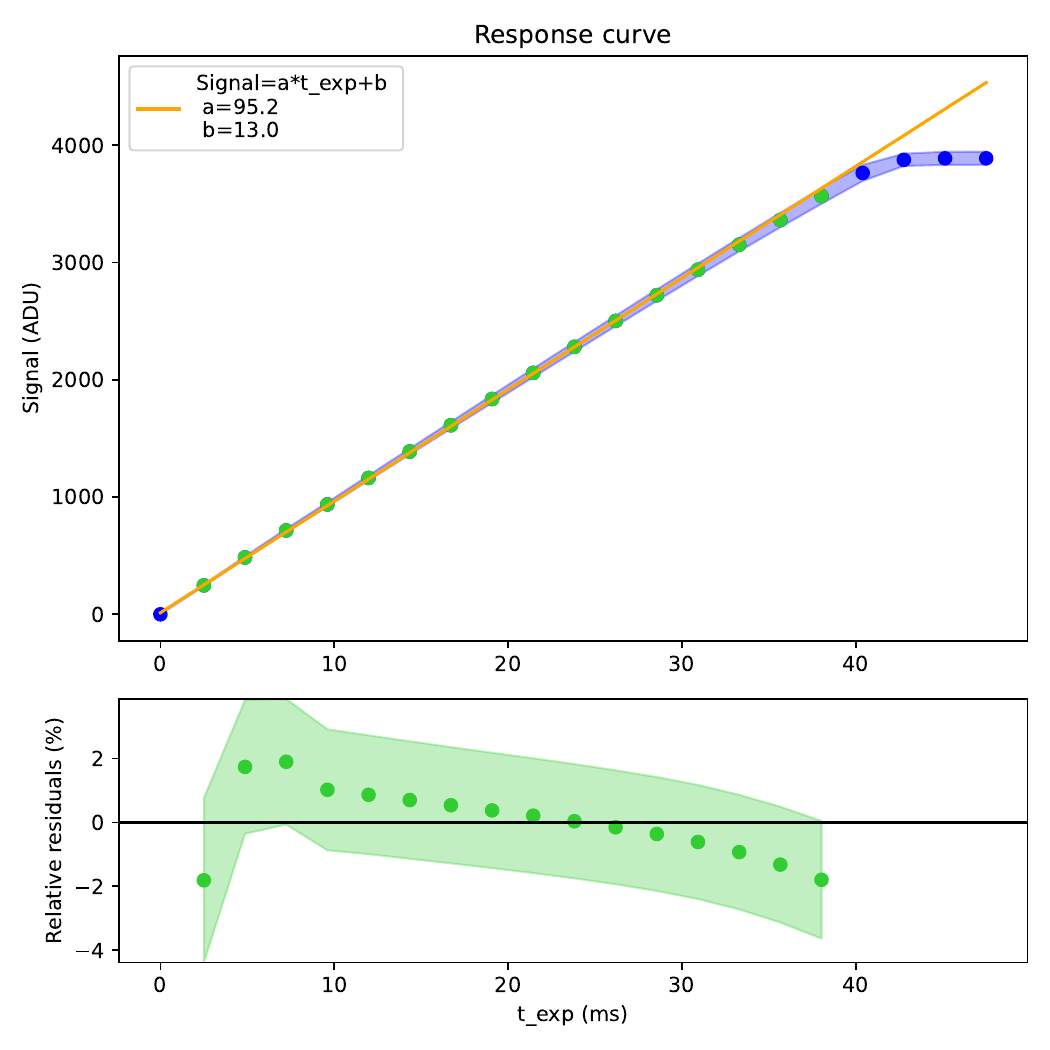}
    \caption{The top graph shows the response curve, that means of the measured signal over the detector, $\mu_F$, versus the exposure time (green points represent the used data, and blue points represent the unused data). The blue area represents the $1\sigma$ deviation, and the orange line refers to the mean linear model. The relative residuals are plotted below.}
    \label{fig:Response_curve}
\end{figure}

\begin{figure}
    \centering
    \includegraphics[width=0.8\linewidth]{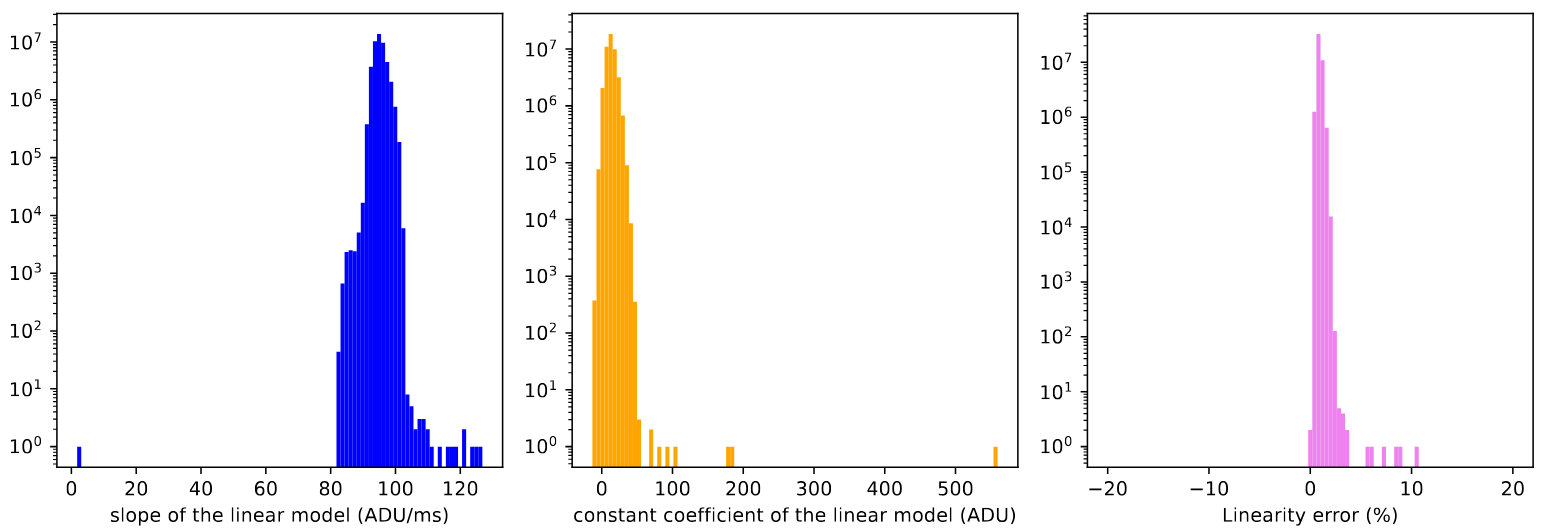}
    \caption{The histograms from left to right depict the slope and constant coefficient of the linear model of the response curve (Fig.~\ref{fig:Response_curve}), as well as the linearity error for each pixel.}
    \label{fig:Linearity}
\end{figure}

The flat signal, corrected for the master dark and master bias, is theoretically linear with exposure time before saturation (approximately 5\,000\,e$^-$ in high-gain mode and 50\,000\,e$^-$ in low-gain mode). This allows the linearity error (LE) of the detector to be calculated as:
\begin{equation}
    \textrm{LE}=100\frac{1}{n}\sum_{i=1}^{n}\frac{\mid \mu_{\mathrm{F},i}-(a t_i+b)\mid}{a t_i+b},
    \label{eq2}
\end{equation}
where $n$ is the number of measurements, $\mu_{\mathrm{F},i}$ the measured flat signal during $t_i$ the exposure time of the acquisition, $a$ and $b$ the parameters of the linear model. We measured a mean $\mathrm{LE}=0.90\pm0.18\,\%$ in low-gain mode. This result is based on pixel’s filling between 5\% and 95\% of the full well dynamic. The response curve and its residuals are plotted in Fig.~\ref{fig:Response_curve}. The curve exhibits the classic linearity behavior of detectors with an asymptotic shape close to saturation. The high relative non-linearity at low filling probably comes from the low denominator values used to compute the relative error. Figure~\ref{fig:Linearity} shows the histograms of the slope and the constant coefficient of the linear model of the response curve, as well as the linearity error for each pixel.


\subsection{Readout and Quantization Noise}
\label{sec:roq_noise}

\begin{figure}
    \centering
    \includegraphics[width=1\linewidth]{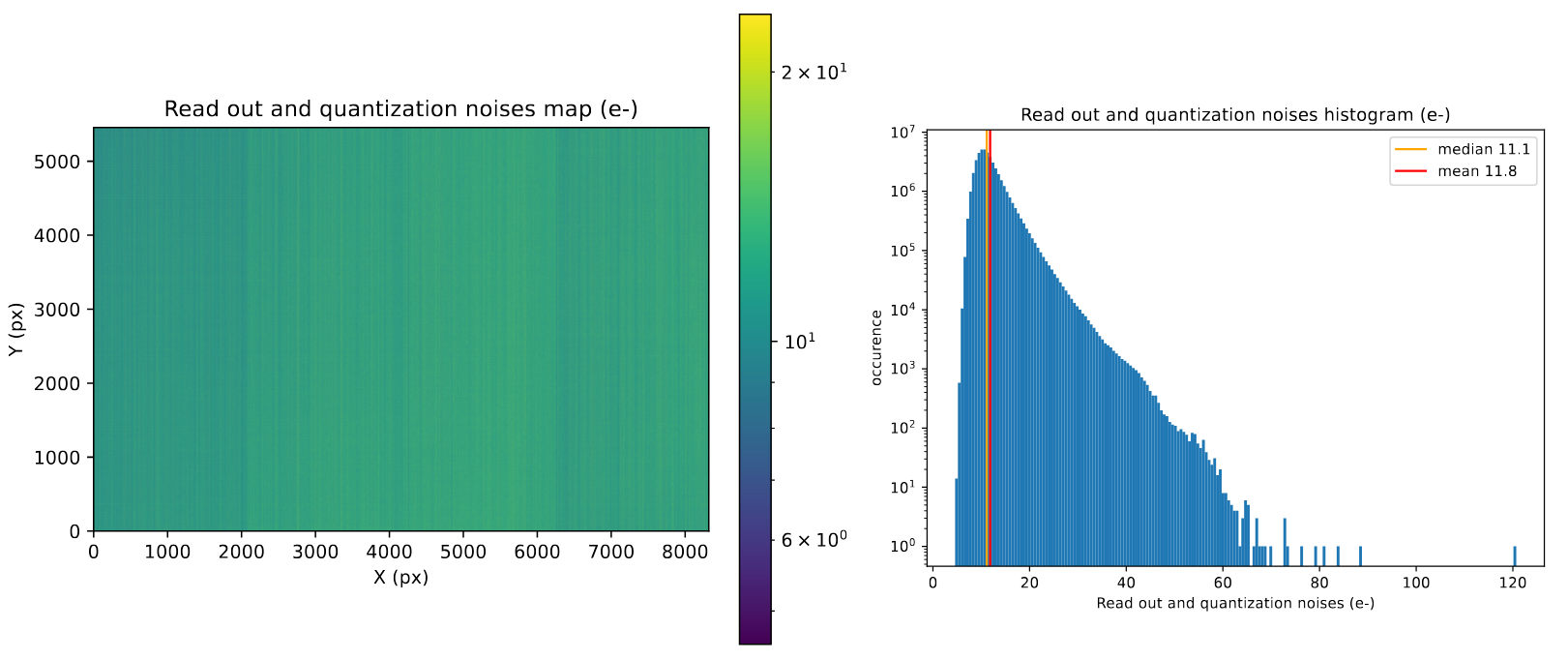}
    \caption{Read-out and quantization noise over the detector, map (left) and histogram (right).}
    \label{fig:ROQ_noise}
\end{figure}

The measured bias images $\mu_{\mathrm{B}\,i}$ are subtracted in pairs and the variance is used to compute the readout and quantization noise ($\sigma_{\mathrm{roq}}$):
\begin{equation}
    \sigma^2_{\mathrm {roq}} = \frac{\mathrm{Var}_i(\mu_{\mathrm{B\ i}}-\mu_{\mathrm{B\ i+1}})}{2},
    \label{eq:roq_noise}
\end{equation}

The readout and quantization noise over the detector is shown in Fig.~\ref{fig:ROQ_noise}, the mean value is $12\pm3$\,e$^-$. The manufacturer Pyxalis measured $16\pm3$\,e$^-$, this difference may be explained by various factors: noise depends on the frame rate and they may not have used the same one as us. Besides, they compute the gain factor on the beginning of the filling ramp, while we use the full dynamic, thus they estimate an ADU-to-electron conversion that is smaller than ours, which leads to higher noise values in electrons. Eventually, they use a dedicated proximity electronics (not the GENEPYX camera, which might lead to excess noise. Fig.~\ref{fig:ROQ_noise} exhibits a column behavior of the noise which might come from the ADC at the upper and bottom parts of the matrix.

We can calculate the quantization noise~\cite{Healey_1994} as $\sigma_q=q/\sqrt{12}$, where $q$ is the quantization step value estimated in Sec.~\ref{sec:PTC}. With $q=13.9$~e$^-$, therefore the quantization noise is $\sigma_q=4.0$~e$^-$.  The readout noise is then given by:  $\sigma_{ro}=\sqrt{\sigma_{roq}^2-\sigma_q^2}=11.3$~e$^-$.

\subsection{Dark Current}
\label{sec:dark_current}

\begin{figure}
    \centering
    \includegraphics[width=1\linewidth]{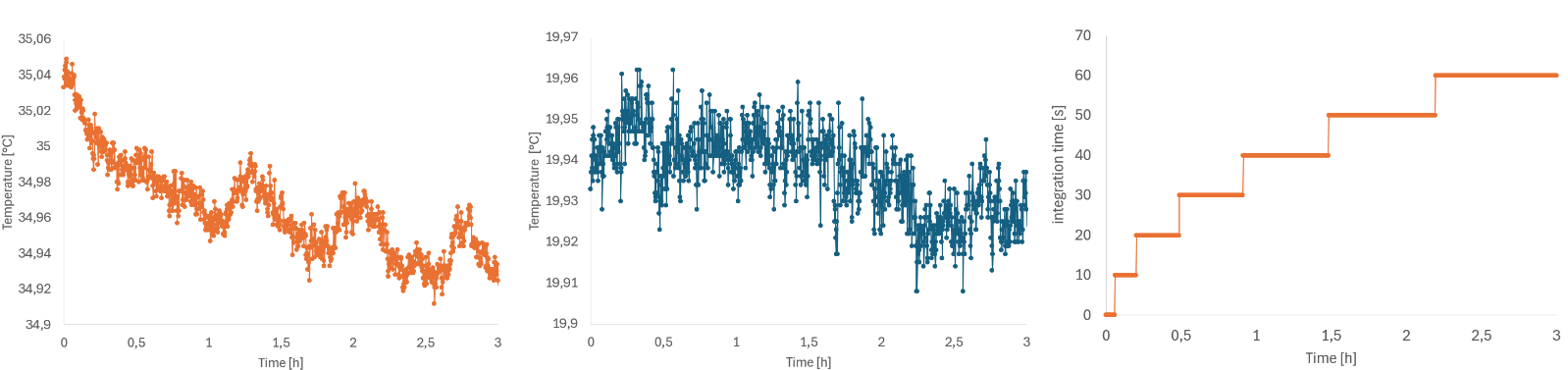}
    \caption{Left: evolution of the detector temperature during the acquisition. Center: evolution of the room temperature during the acquisition. Right: evolution of the exposure time during the acquisition.}
    \label{fig:temperature}
\end{figure}

Dark current measurements are highly sensitive to temperature. After completing the initialization stage, we adjust the exposure time for our acquisitions to capture various signal levels. This results in variations in the detector's readout frequency, that affect temperature. Therefore, we adapt the frame rate to minimize temperature variation throughout the entire integration time. The measurements are presented in Fig.~\ref{fig:temperature}. We observe that the room temperature remains stable at $0.05^\circ$C. The detector temperature is also quite stable at approximately $35^\circ$C, varying by less than $0.1^\circ$C. However, as previously explained a slight decrease is observed in correlation with an increase in exposure time.  This temperature change is expected to result in a dark current change of less than $0.3$\,e$^-$/s based on the dark current doubling factor of $7.2^\circ$C given by the manufacturer~\cite{Michelot_2025}. This dark current variation will have a minimal impact on the linear regression used to obtain dark current values.

\begin{figure}
    \centering
    \includegraphics[width=0.7\linewidth]{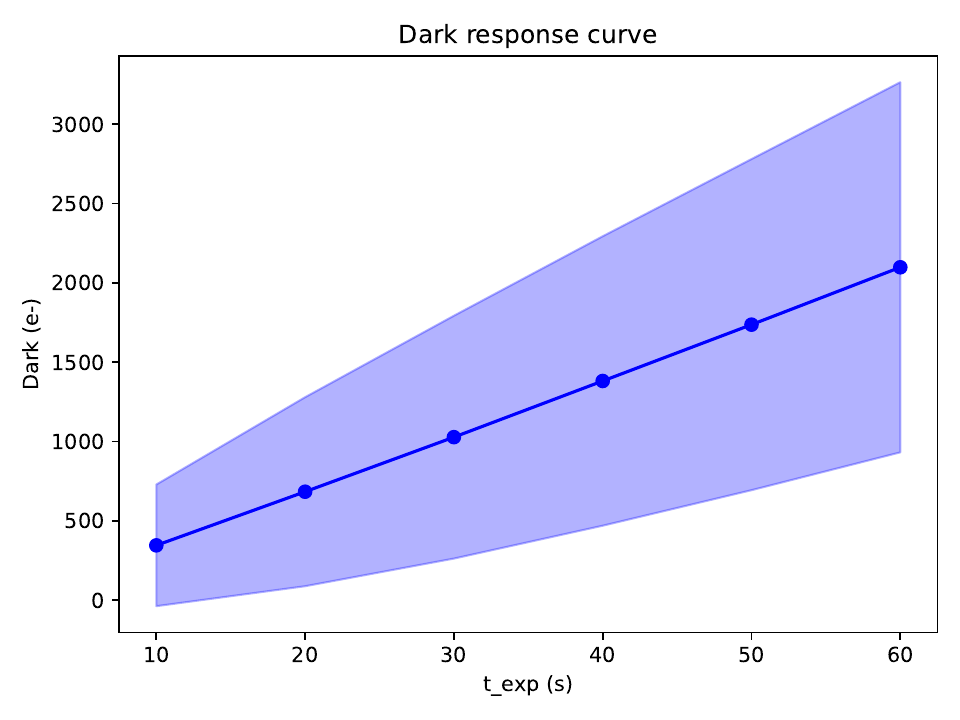}
    \caption{Dark current integration curve. The solid line that matches the circle markers shows the mean dark current $\mu_D$} over the detector versus exposure time. The blue area represents the $1\sigma$ deviation.
    \label{fig:dark_response_median}
\end{figure}
The dark signal is linear in exposure time and the slope gives the dark current, this curve is presented in Fig.~\ref{fig:dark_response_median}. We found a median value over the detector of $35\pm18$\,e$^-$/s at $35^\circ$C.

\subsection{Pixel response non uniformity (PRNU) and dark signal non uniformity (DSNU)}
\label{sec:PRNU}

The PRNU is calculated by subtracting the spatial variances ($\sigma_F^2$ and $\sigma_D^2$) and the spatial mean values ($\mu_F$ and $\mu_D$) of the flat and dark frames with no illumination and at 50\% of the detector saturation (corresponding roughly to $25$~ms):
\begin{equation}
    PRNU=100\frac{\sqrt{\sigma_{F\ 50\%}^2-\sigma_{\mathrm{D}}^2}}{\mu_{F\ 50\%}-\mu_{\mathrm{D}}}
    \label{eq3}
\end{equation}

\begin{figure}
    \centering
    \includegraphics[width=0.9\linewidth]{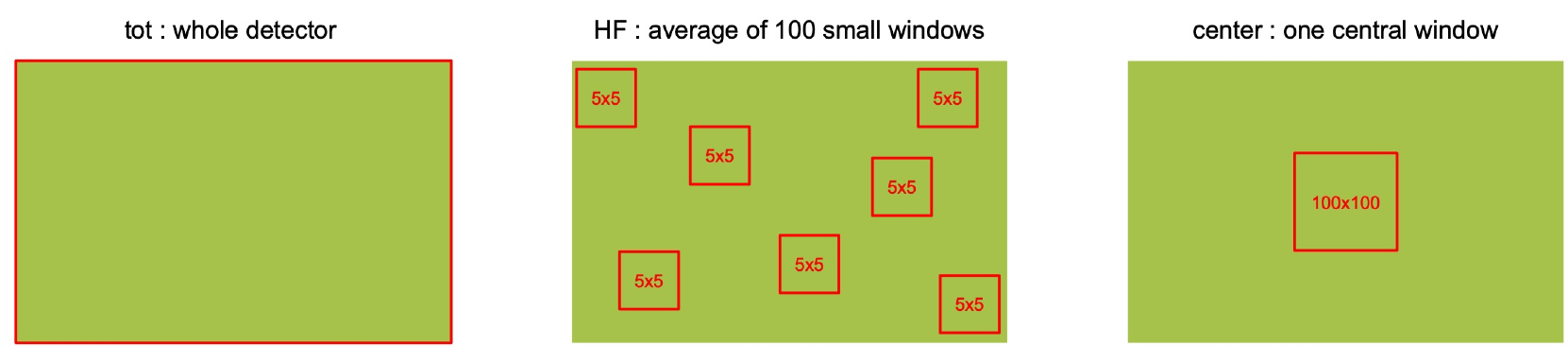}
    \caption{Positions of the different areas to calculate different values of the PRNU: over the whole detector (left), averaged using 100 small windows (center) and computed on a centered window.}
    \label{fig:PRNU}
\end{figure}

We defined 3 different PRNU : the total PRNU$_\mathrm{tot}$ over the whole detector, the high frequency PRNU$_\mathrm{HF}$ which is an average over 100  of $5\times5$ pixels windows, and the PRNU$_\mathrm{center}$ over the $100\times100$ centered window. Their geometries are illustrated in Fig.~\ref{fig:PRNU}. We found PRNU$_\mathrm{tot}=1.8\,\%$, PRNU$_\mathrm{HF}=1.1\,\%$, PRNU$_\mathrm{center}=1.3\,\%$.

The DSNU was also investigated using the master dark with an exposure time of $6$~s which is illustrated in Fig.~\ref{fig:acquired_images} We found DSNU$_\mathrm{tot}=1384~\mathrm{e}^-$, DSNU$_\mathrm{HF}=1340~\mathrm{e}^-$, DSNU$_\mathrm{center}=696~\mathrm{e}^-$.

\subsection{Defective pixels}
\label{sec:defect_px}

Bad pixels can introduce false detection results and reduce the accuracy of photometry and astrometry. A defective pixel is defined as an anomalous~\cite{An_2007, Tchendjou_2021, Zhao_2024, Jia_2024} pixel, which means that defective pixels are different from the rest of the pixels. Although there is no exact definition of a defective pixel, it depends on the required application. To illustrate this diversity, we can examine various definitions proposed in the literature. Zhao et al.~\cite{Zhao_2024} proposed two criteria to distinguish between flickering pixels and non-linear response pixels. Hu et al.~\cite{hu2022sdgsat} defined two ways to identify non-effective pixels: the first by studying the relationship between responsiveness and radiant temperature, and the second by studying responsiveness at a given radiant temperature. Wang et al.~\cite{wang2019new} chose a simple criterion to distinguish pixels that were obviously higher or lower than the other pixel values. Jing et al.~\cite{jing2013research} distinguished cold, hot, and non-saturated pixels by selecting an appropriate threshold value.

\begin{figure}
    \centering
    \includegraphics[width=1\linewidth]{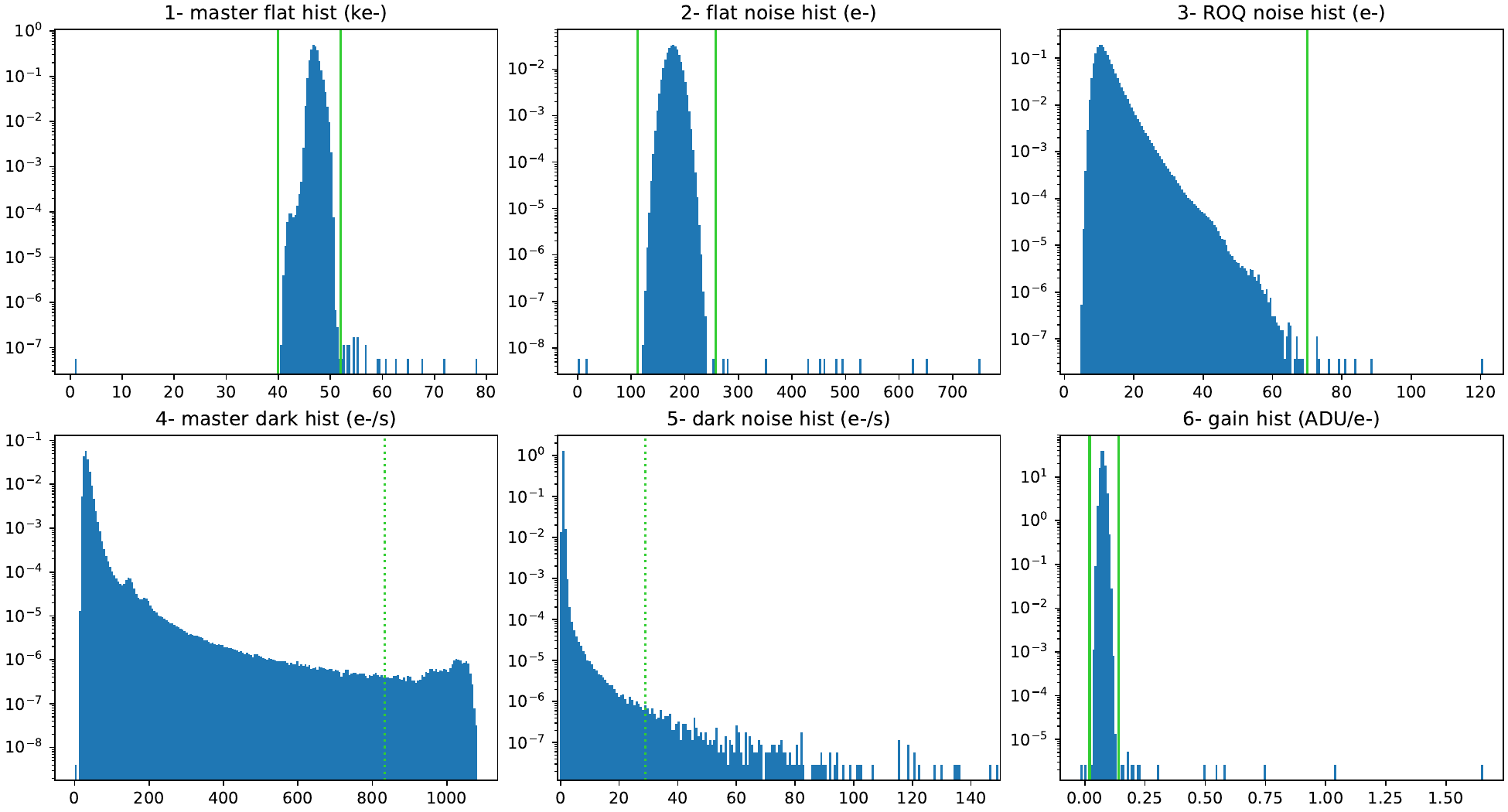}
    \caption{Identification of defective pixels using histograms with green thresholds to distinguish each type. The top left corresponds to hot and cold pixels, the top center to noisy response pixels, the top right to noisy readout pixels, the bottom left to anomalous dark current pixels, the bottom center to noisy dark current pixels, and the bottom right to anomalous gain pixels. The solid lines correspond to the thresholds used to identify these outlier pixels (see ~\ref{sec:defect_px}). The dotted lines correspond to the dark current and dark current variation thresholds that exceed the specific values related to the science case (see the discussion in ~\ref{sec:defective_pixels_2}).}
    \label{fig:defect_px_hist_mosaic}
\end{figure}

In this paper we consider pixels that are outliers to be defective. To identify and discriminate problematic pixels, we categorized six types of statistical behavior, see Table~\ref{tab:defect_px}. Their statistics are displayed in Fig.~\ref{fig:defect_px_hist_mosaic} and the thresholds used to identify outliers are shown as solid lines.
\begin{table}
\caption{Description and number of each type of defective pixel} 
\label{tab:defect_px}
\begin{center}       
\begin{tabular}{llll}
\hline
\textbf{Defect type} &\textbf{Description} &\textbf{Nature of defect} &\textbf{Number of pixels}\\
\hline
\hline
1 &Hot and cold pixels &spatial &24\\
2 &Noisy response pixels &temporal &14\\
3 &Noisy readout pixels &temporal &10\\
4 &Anomalous dark current pixels &spatial &0\\
5 &Noisy dark current pixels &temporal  &0\\
6 &Anomalous gains pixels &spatial  &17\\
\hline
\textbf{Total} & & &37\\
\hline
\end{tabular}
\end{center}
\end{table} 
\begin{figure}
    \centering
    \includegraphics[width=0.6\linewidth]{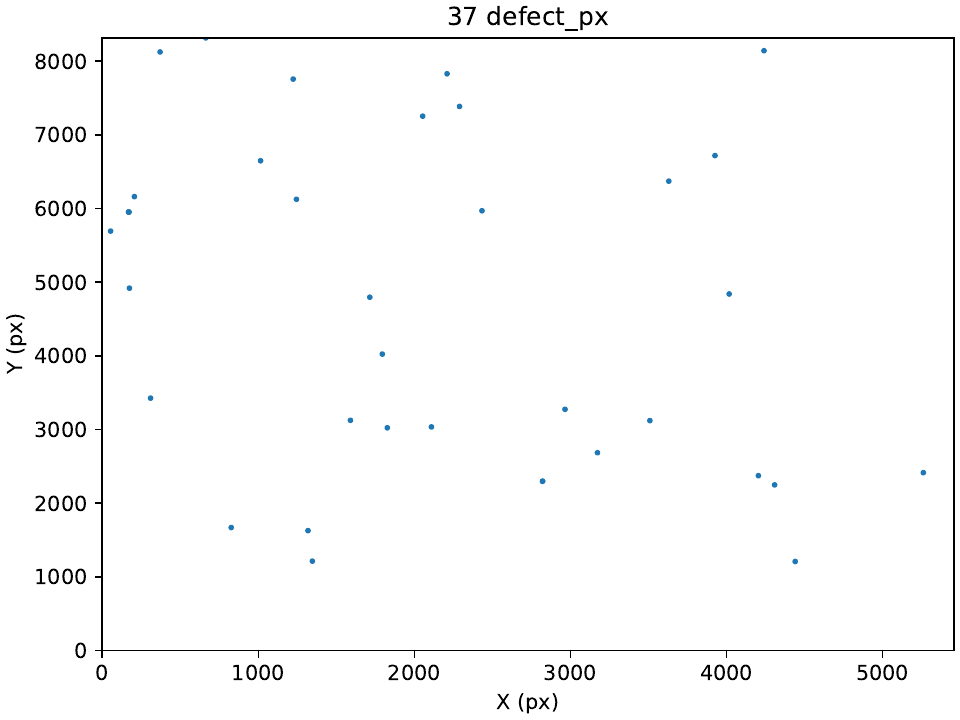}
    \caption{Map of the defective pixels over the detector.}
    \label{fig:defect_px_map}
\end{figure}
\begin{enumerate}
    \item Hot and cold pixels: pixels with an unacceptable response to a photon flux (too high or too low).
    \item Noisy response pixels: pixels with noisy temporal response, meaning flashing or flickering pixels.
    \item Noisy readout pixels: pixels with an excess of readout noise.
    \item Anomalous dark current pixels: pixels with a high dark current. There are no outliers present in our case.
    \item Noisy dark current: pixels with an excess of temporal dark noise. There are no outliers present in our case.
    \item Anomalous gains pixels: pixels with anomalous gains.
\end{enumerate}
Thanks to the above-described criteria, we can identify the defective pixels. Figure~\ref{fig:defect_px_map}, and Table~\ref{tab:defect_px} details their number as a function of type. A total of $37$ defective pixels were found across the detector, corresponding to $8\times10^{-5}\%$ of the total number of pixels.

\subsection{Inter-Pixel Capacitance}
\label{sec:IPC}

\begin{figure}
    \centering
    \includegraphics[width=0.6\linewidth]{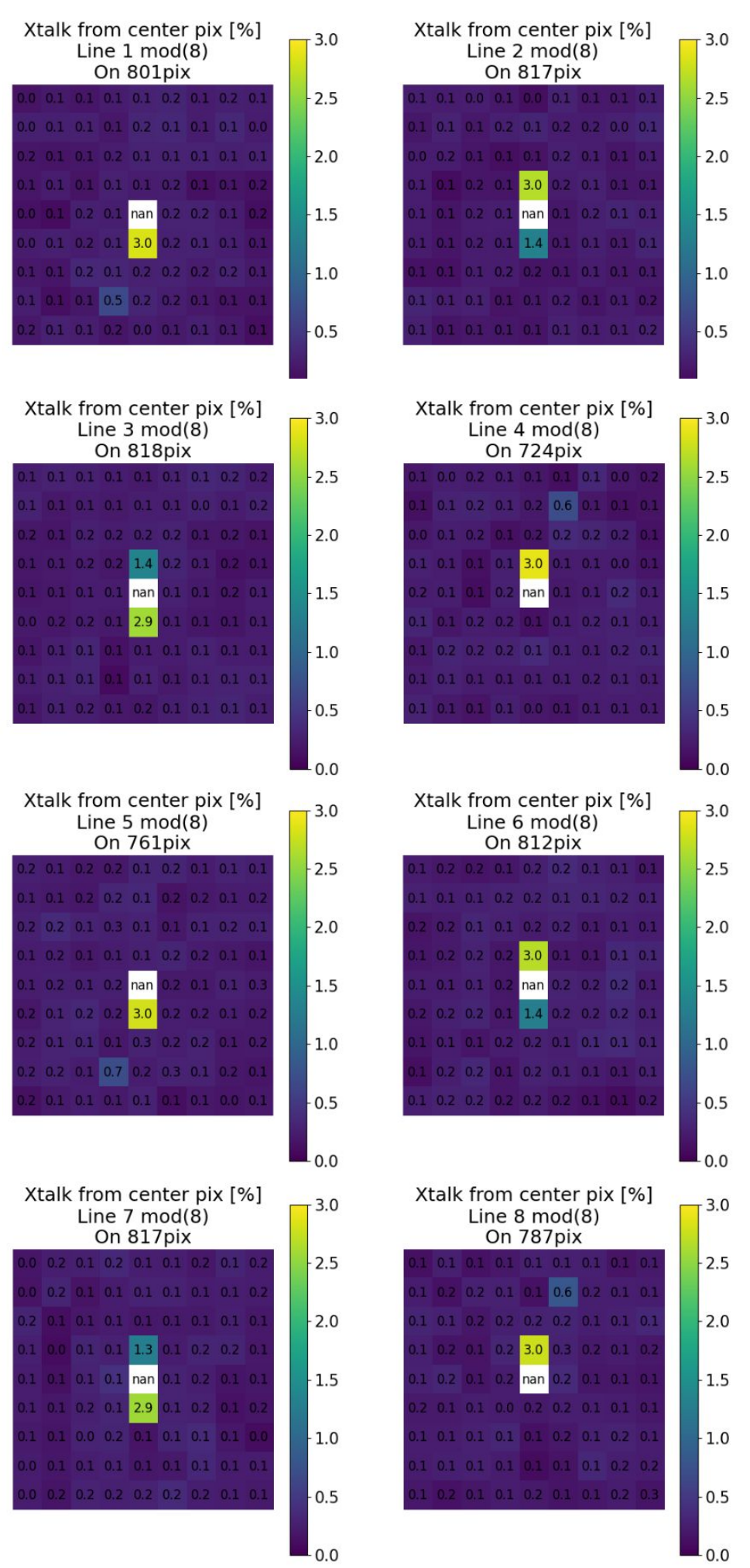}
    \caption{Inter-pixel capacitance maps in \% of the central pixel value for the modulo eight lines on the whole detector. Center white pixel is the hot/influencing one. The neighbors are not-warm pixels.}
    \label{fig:crosstalk_map}
\end{figure}
The cross-talk is a metric that quantifies the impact of the value of a single pixel on the values of its neighboring pixels. This parameter is of primary importance for missions with high accuracy requirements and significant illumination gradients. This principle applies in particular to astrometric missions that entail the observation and precise determination of stellar centroids.

Cross-talk can emanate from two distinct sources within the detector. The first one is the cross-talk attributable to the detection layer, induced by the migration of photocarriers (electrons or holes) from the area of one pixel to its neighbors prior to its collection by the photodiode and integration for signal contribution. A priori, the Gigapyx detector is not capable of exhibiting this type of cross-talk due to the presence of deep trenches~\cite{Ahmed_2014} in the silicon detection layer, which prevent carriers from migrating from one pixel to another. This assertion necessitates validation through dedicated setups that actually generate free carriers in the detection layer, like single pixel illumination ~\cite{layden2025characterizationteledynecosmoscamera}, utilization of \textsuperscript{55}Fe sources or cosmic ray impact studies~\cite{Le_Graet_2022}.

The second source of cross-talk is named the IPC and is attributed to capacitive effects between the photodiodes and in the readout circuits. The proximity of the readout electrical lines, in conjunction with a high readout frequency, induces a finite impedance between two pixel circuits, resulting in one pixel impacting another. This effect is well described in Ref~\cite{Moore_2006}.

In this particular context, the decision was made to characterize the Gigapyx inter-pixel capacitance utilizing the hot pixels of the matrix~\cite{Le_Graet_2022,Donlon_2018}. Indeed, given our emphasis on the analysis of inter-pixel capacitance in readout circuits, it is essential that we consider the ideal scenario in which a high integrated signal pixel is surrounded by low integrated signal pixels. This ideal case is analogous to a hot pixel during a dark-acquisition process. What we call "hot pixels" in this section are pixels with substantial higher than average values. To ensure the integrity of the study's findings, the effects of hot pixels with hot neighbors are mitigated through a rigorous process of exclusion. Indeed, the $9\times9$ windows considered for the study are taken around a central pixel with a higher than average value surrounded with average pixels (i.e. with values around the average of the expected dark signal).

The measured inter-pixel capacitance pattern exhibits an 8-line repeatability. It has been observed that each pixel has an effect on two adjacent pixels. The impacted neighbors are the ones above and under for lines 2, 3, 6 and 7 modulo 8. The impacted neighbors are identified as those located above, below, and a separated one three lines away on a different column for the lines 1, 4, 5, and 8 modulo 8. Following consultations with the manufacturer, Pyxalis, it was determined that this is in accordance with the detector's readout lines layout. As illustrated in Fig.~\ref{fig:crosstalk_map}, the line-dependent kernel is averaged across the entire matrix for the modulo 8 lines.

\begin{figure}
    \centering
    \includegraphics[width=1\linewidth]{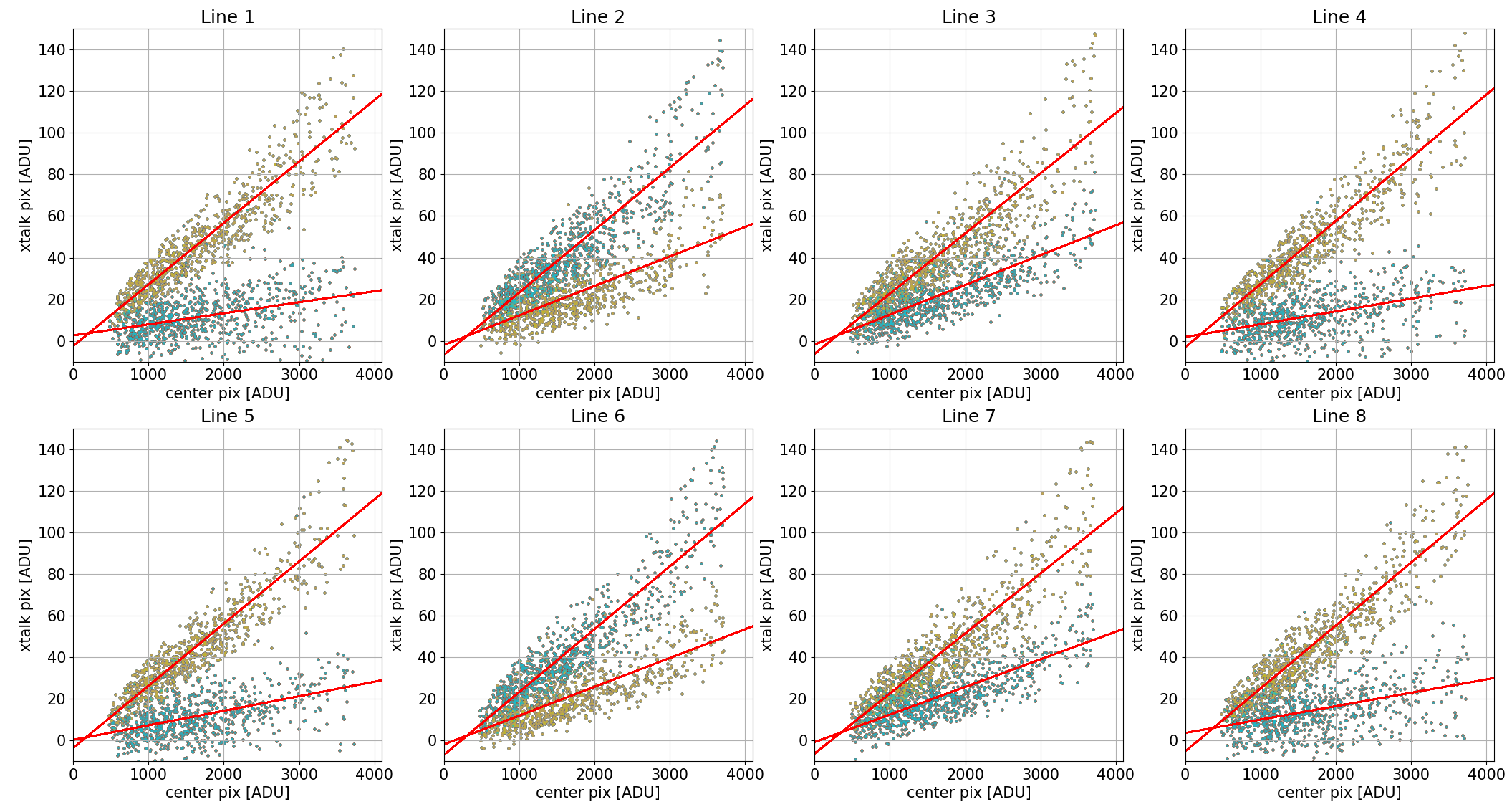}
    \caption{Inter-pixel capacitance linear regressions on the neighbor pixels values with regard to central pixel value. Only the two principal influenced neighbor pixels are represented}
    \label{fig:crosstalk_regressions}
\end{figure}
The inter-pixel capacitance maps were computed based on 50 acquisitions of $10$\,s integration time each, which were then averaged to create a master image. In the analysis of the master image, the pixels that exhibit higher values in comparison to the mean matrix value are identified (i.e., "warm pixels"). Additionally, the absence of "warm neighbors" within a radius of up to four pixels is determined. The mean of the $9\times9$ windows surrounding the identified hot pixels is calculated by modulo 8 lines. It is estimated that approximately $1400$ pixels are incorporated into each final kernel. Subsequently, a linear regression analysis is performed on the neighbor pixel values relative to the central pixel value, thereby quantifying the impact of the central pixel on its neighboring pixels, as illustrated in Fig.~\ref{fig:crosstalk_regressions}.

It is observed that a central pixel exhibits inter-pixel capacitance with two adjacent pixels. The impact on the neighbor pixel is proportional to the central pixel value. In the most unfavorable circumstances, the inter-pixel capacitance levels are less than 3$\%$ in the study configuration. These levels are excessive for astrometry scientific requirements, but can be mitigated by simple post-processing techniques accounting for spatially varying kernels (cf.\ ~\ref{sec:IPC-correction}). Subsequent steps for us will be to evaluate the impact of IPC (before and after post-processing) on the estimation of the parameters of interest, that is to say the positions of the observed sources. Preliminary tests have indicated that adjusting the detector configuration could reduce inter-pixel capacitance levels. However, this would result in a decrease in the component's frame rate.

\section{Discussion}
\label{sec:discussion}

The performance of the GIGAPYX-4600 detector and therefore of the Gigapyx technology must be discussed in the context of astrometry since the objective is to develop a focal plane based on this technology~\cite{Amiaux_2025}. This study was carried out independently from the Pyxalis manufacturer. Therefore, it is worthwhile to compare first our results with the data sheet (Sec.~\ref{sec:comparison}). Then, we can address the expected performance in space conditions (Sec.~\ref{sec:space}).

\subsection{IPAG/Pyxalis comparison}
\label{sec:comparison}

\begin{table}
\caption{46M Gigapyx characterization, comparison between IPAG and Pyxalis} 
\label{tab:results_comparison}
\begin{center}  
\begin{tabular}{lll}
\hline
Property & Pyxalis & IPAG\\
\hline\hline
 read out noise & 16 e$^-$ & 12 e$^-$\\
 linearity error & 2\% & 0.90\%\\
 response non uniformity & $\leq$ 2\% & 1.3\%\\
 dark current (at $35^\circ$C) & 24 e$^-$/s & 35 e$^-$/s\\
 saturation charge & $>$50\,ke$^-$ & 52\,ke$^-$\\
 percentage of defective pixels &$\leq$ 0.1\%& $8\times10^{-5}\%$ \\
 inter-pixel capacitance &- & $\leq$ 3\% \\ 
\hline
\end{tabular}
\end{center}
\end{table} 

Table~\ref{tab:results_comparison} shows a comparison of the results of our investigation with the Pyxalis results given in their datasheet. Our results are very close to the claims in the data sheet, except for the dark current value.

We measured a median mean dark current value of $35\pm18$\,e$^-$/s over the detector at $35^\circ$C whereas Pyxalis measured~\cite{Michelot_2025} $24$\,e$^-$/s. Our measurements are slightly higher than the Pyxalis data. After discussing this issue with the manufacturer, we learned that they are working on an improvement that may result in significantly lower dark currents in future versions of this detector.

\subsection{Effect on centroid measurement accuracy}
\label{sec:centroid}

Dark current, linearity, bad pixels and cross-talk can reduce the accuracy of photometry and astrometry. A number of effects can produce systematic errors in centroid measurements. Since an illumination correction can be performed, many non-uniformities that are not wavelength dependent can be calibrated out and not contribute systematic error. However it is not the purpose of this paper to investigate the effects that has been measured on the accuracy of the measured centroids.

\subsection{Defective pixels}
\label{sec:defective_pixels_2}

A total of $37$ outliers pixels were found across the detector in Sec.~\ref{sec:defect_px}. Although, this value is entirely driven by the requirements of the end-user mission because, there is no exact definition for one defective pixel. Below, we can see a definition that can be more realistic in the context of astrometric space missions.

\begin{enumerate}
    \item Hot and cold pixels: this aspect is not very constraining in the context of astronomical observations because the final images will be flat-corrected. Therefore, the criterion described in ~\ref{sec:defect_px} can be kept.
    \item Noisy response pixels: flashing pixels must be removed. The criterion described in ~\ref{sec:defect_px} can be kept.
    \item Noisy readout pixels: the threshold can be set quite high since the observation strategy involves integrating strong light fluxes, leading to the photon noise regime. Therefore, a weight can be assigned to noisy pixels to reduce their influence.  The criterion described in ~\ref{sec:defect_px} can be kept.
    \item Anomalous dark current pixels: pixels with a high dark current. If we look for signals higher than $5\%$ of the full well being occupied by dark electrons during a $2$\,s exposure (the maximum exposure time with HWO) at a temperature of 290\,K (the HWO temperature), then it corresponds to signals greater than $833$\,e$^-$/s. However this criterion is not very important because the final images will be dark-corrected.
    \item Noisy dark current: pixels that can be defined with a dark current variation greater than the noise corresponding to the limit described in type\,4 ($\sqrt{833}=29$\,e$^-$/s).
    \item Anomalous gains pixels: for the same reason as with hot and cold pixels, the criterion described in ~\ref{sec:defect_px} can be used.
\end{enumerate}

The criteria 1, 2, 3 and 6 are the same as defined in Sec.~\ref{sec:defect_px}, and criteria 4 and 5 are illustrated as dotted lines in Fig.~\ref{fig:defect_px_hist_mosaic}.

Thanks to the above-described criteria we can identify the defective pixels. A total of $6471$ defective pixels were found across the detector, corresponding to $0.014\%$ of the total number of pixels.

\subsection{Correction of Inter-Pixel Capacitance}
\label{sec:IPC-correction}

\begin{figure}
    \centering
    \includegraphics[width=\textwidth]{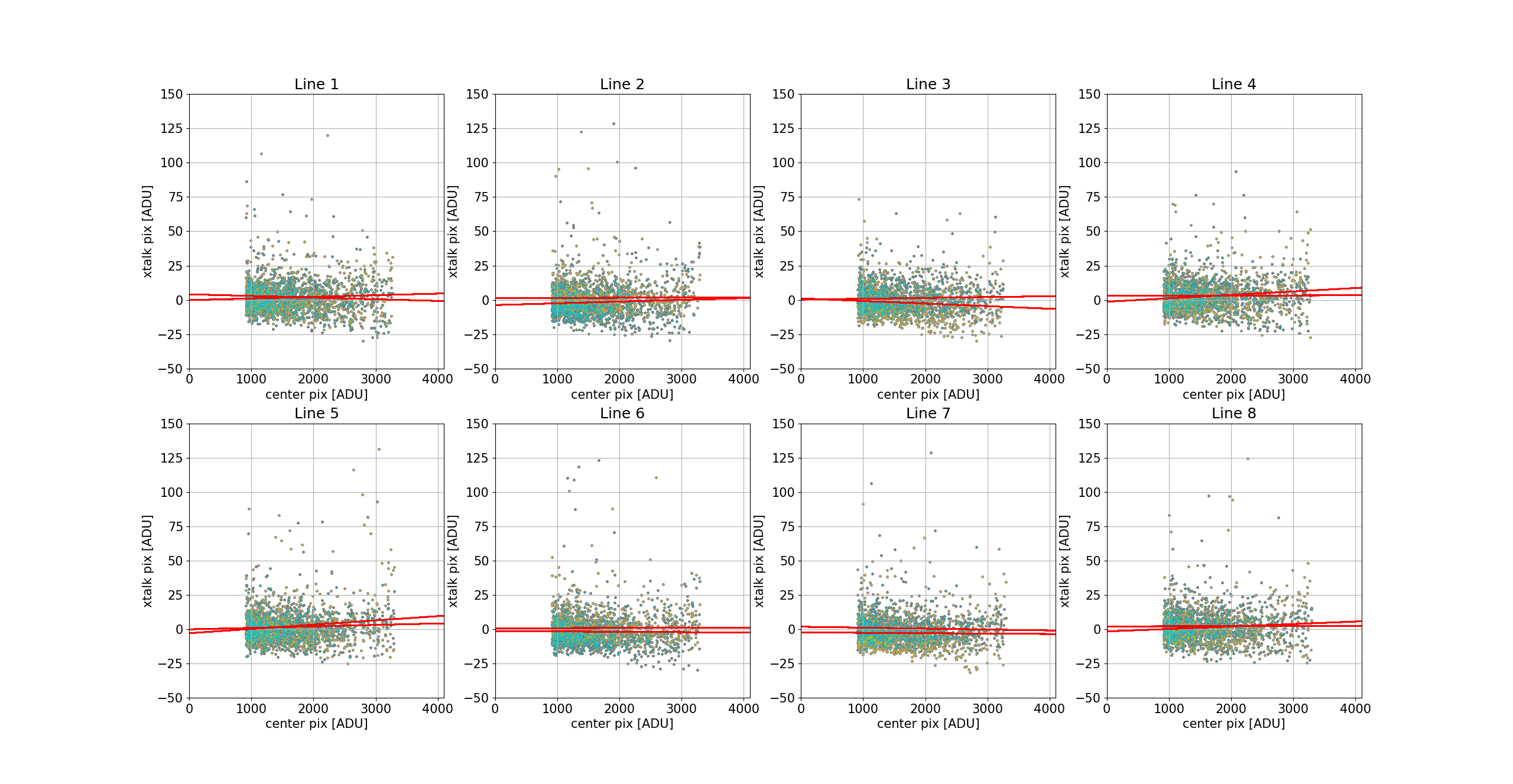}
    \caption{After correction of the inter-pixel capacitance, linear regressions on the neighbor pixels values with regard to central pixel value. Only the two principal influenced neighbor pixels are represented}
    \label{fig:deconv_crosstalk_regressions}
\end{figure}

For high resolution astrometry requirements, we need to cancel out the IPC of the images. We cannot deconvolve the IPC signal presented earlier with simple algorithms like the Wiener filter which is a simple frequency-wise multiplication in the Fourier space since the kernel is spatially varying. Two algorithms that can deal with this issue were studies in this paper. The first one is adapted from the algorithm proposed to compensate IPC for the infrared H2RG arrays of the James Webb Space Telescope~\cite{Donlon_2018}. This algorithm is an iterative process based on the fact that the IPC coefficients are sufficiently low to allow to successively approximate the real signal by the measured one and eventually converge to the real signal without IPC contribution. In Appendix~\ref{sec:conjugate-gradient-algorithm}, we propose another method based on a regularized weighted least squares fit of the raw image $\mu$ which, owing to the size of the problem ($\sim 46$ millions pixels), is solved by means of the iterative method of conjugate gradients~\cite{Hestenes_Stiefel-1952-conjugate_gradients}. The two algorithms find the same solution in a few seconds of computations, though the conjugate gradients method offers well studied guaranties of convergence which is not the case of the other method. A first result of correction using the algorithm described in Appendix~\ref{sec:conjugate-gradient-algorithm} is presented in Fig.~\ref{fig:deconv_crosstalk_regressions} and can be compared to Fig.~\ref{fig:crosstalk_regressions}.

\subsection{Performances in space environment}
\label{sec:space}

Pyxalis evaluated the performance~\cite{Michelot_2025} of the Gigapyx in proton and heavy ion environments to assess its radiation hardness. These tests were carried out after proton irradiations up to a total fluence of $2.3 \times 10^{11}$~p+/cm$^2$ (62 MeV). The degradation mostly impacts the dark current level, the dark current heterogeneity and the temporal noise. The other performances such as non-linearity, saturation charge and pixel response non-uniformity, remain unchanged. These degradation orders of magnitude are consistent with those of similar technologies that have already been published.

Results obtained from heavy ion irradiations demonstrate that the GIGAPYX-4600 sensor is latch-up free at least up to $57\,\mathrm{MeV} . \mathrm{cm}^2 / \mathrm{mg}$, and resistant to the blooming effects induced by single-event transients at pixel level, thanks to its capacitive deep trench isolation. Various single-event effects have been studied and yielded promising results for a device of this kind operating in space. These radiations results will be used as inputs when designing space cameras based on the Gigapyx sensors. In the near future, Pyxalis will provide off-the-shelf radiation-hardened readout electronics for this sensor family.

\subsection{Impact on high precision astrometry}

High-precision astrometry places strict requirements on detector characterization, especially for missions such as HWO and Theia, where accurate centroiding at the sub-microarcsecond level is critical. This study thoroughly evaluates the GIGAPYX-4600’s noise, linearity, dark current, and inter-pixel capacitance. However, several key aspects that directly influence centroid measurement precision remain to be addressed.

First, pixel response non-uniformity at sub-percent levels must be calibrated to avoid bias in stellar position measurements, particularly for faint sources where photon statistics dominate. Second, the temporal stability of the detector’s response (e.g., gain drift and dark current evolution) over long exposure times or varying thermal conditions needs further investigation because such variations could induce systematic shifts in centroid positions.

Additionally, this study did not assess geometric distortions like pixel scale variations and orthogonality errors, or optical aberrations like field-dependent variations of the point- spread function (PSF), both of which are essential for astrometric precision. These effects require dedicated calibration using laboratory metrology~\cite{Crouzier_2016, Shao_2023} and star fields~\cite{Malbet_2024, Lizzana_2025}. Finally, stray light and persistence effects, which are common in space environments, could degrade centroiding accuracy if not mitigated.

Future work should include end-to-end simulations that combine detector characterization data with realistic PSF models to quantify how these unaddressed factors propagate into centroiding errors. These simulations would ensure that the GIGAPYX-4600 can measure 0.3\,\textmu{}as signals, which are required for next-generation astrometry missions.

\section{Conclusion}
\label{sec:conclusion}

In the context of space-borne astrometric observations and the development of next-generation telescopes, we conducted a comprehensive characterization of the GIGAPYX-4600 CMOS detector manufactured by Pyxalis at IPAG. Our results confirm that the detector meets the manufacturer’s performance specifications, with linearity errors below 1\%, pixel response non-uniformity $\leq1.3\%$, readout noise of $12\pm3\,\mathrm{e}^{-}$, and inter-pixel capacitance under 3\%. The dark current of $35\pm18\,\mathrm{e}^{-}/\mathrm{s}$ at $35^\circ\mathrm{C}$ and the extremely low defective pixel rate further validate its suitability for high-precision applications. Additionally, we demonstrated that inter-pixel capacitance effects can be effectively mitigated by post-processing images with dedicated spatially varying kernel correction algorithms, ensuring compatibility with high-precision astrometry requirements. Radiation tests also demonstrated its robustness against protons and heavy ions, with no latch-up observed, making it a strong candidate for space missions.

The GIGAPYX-4600 stands out as a significant proof of concept for large-format CMOS detectors in astrometry, particularly for missions like HWO and Theia, which require gigapixel-scale focal planes with microarcsecond-level precision. Given its scalability, with larger Gigapyx detectors up to 220~Mpx, the GIGAPYX-4600 paves the way for modular focal plane assemblies capable of covering the large fields of view required by HWO and Theia. With ongoing advancements in radiation-hardened readout electronics and crosstalk mitigation, this type of technology is positioned to be key to the next era of high-precision space astrometry.

In the future, we will continue our characterization work on the next format of the Gigapyx family which will be produced, the 220~MP model, which is the largest and which we would like to use in our astrometry missions. The 220~MP detector will be characterized with the standard and the astro-mechanical packages and electronics~\cite{Ardellier-Desages_2026}.

\subsection* {Acknowledgments}

We would like to thank the Pyxalis teams for their support during this project. This work has been partially supported by the LabEx FOCUS ANR-11-LABX-0013 and the CNES agency. The author Manon Lizzana acknowledges financial support from the Centre National d’Études Spatiales (CNES), France (ROR: \url{https://ror.org/04h1h0y33}), and from Pyxalis.

\subsection* {Code and Data Availability}

The code and a representative sample of the data used are available in a Zenodo repository: \url{https://doi.org/10.5281/zenodo.21034923}.

\subsection* {Disclosures}

The authors declare that there are no financial interests, commercial affiliations, or other potential conflicts of interest that could have influenced the objectivity of this research or the writing of this paper.

We collaborated with Pyxalis to understand how the detector works, interpret some results, and compare them with their figures. Pyxalis also provided half of Manon Lizzana's PhD grant. However, the characterization presented in this paper was conducted independently, Pyxalis did not influence the results, and no conflicts of interest were declared.

\appendix

\section{IPC correction by a regularized weighted least squares fit}
\label{sec:conjugate-gradient-algorithm}

To fix the effects of IPC, we propose a method based on a regularized weighted least squares fit of the raw image $\mu \in \mathbb{R}^n$ ($n$ is the number of pixels):
\begin{align}
    \label{eq:IPC-correction}
    \hat{s}
    &= \underset{s \in \mathbb{R}^n}{\text{arg\,min}}
    \Vert H\cdot s - \mu\Vert^2_W + \rho\,\Vert D\cdot s\Vert^2
\end{align}
where $H \in \mathbb{R}^{n \times n}$ is a linear operator modelling the IPC effect on $s \in \mathbb{R}^n$, the signal unaffected by IPC, $W \in \mathbb{R}^{n \times n}$ is a weighting matrix, $D$ is a linear operator (finite differences or identity) to implement the regularization, and $\rho > 0$ is a tuning parameter. The regularization is to avoid amplification of noise in the result~\cite{Thiebaut_2005}. Operator $H$ corresponds to a very sparse matrix: since IPC affects only 3 output pixels per input pixel, the matrix $H$ has only 3 non-zero coefficients per row out of $n$ coefficients. In practice, $H$ can be stored as a general sparse matrix. For our tests, we use another possibility and implemented operators $H$ and $H^{\top}$ as two functions. To account for the noise in the raw image, $W$ should be the precision matrix of the noise, that is the inverse of its covariance matrix. In practice, we consider that $W$ is diagonal with diagonal entries set to zero for defective pixels and to the reciprocal of their estimated variance for the other pixels. Problem~\eqref{eq:IPC-correction} has a closed-form solution $\hat{s} = A^{-1}\cdot b$ with $A = H^{\top}\cdot W\cdot H + \rho\,D^{\top}\cdot D$ and $b = H^{\top}\cdot W\cdot \mu$. Owing to the size of the problem ($n = 5456 \times 8320$), the solution is found by means of the method of conjugate gradients~\cite{Hestenes_Stiefel-1952-conjugate_gradients} which is an iterative method that only requires to provide $b$ and a fast procedure to apply the $n \times n$ operator $A$.

\bibliography{report}   
\bibliographystyle{spiejour}   


\end{document}